\documentclass[aps,prl,reprint,superscriptaddress,nofootinbib]{revtex4-2}

\usepackage{amsmath,amssymb,bm}
\usepackage{graphicx}
\usepackage{xcolor}
\usepackage{hyperref}
\usepackage{microtype}
\usepackage{booktabs}

\hypersetup{colorlinks=true,citecolor=blue,linkcolor=blue,urlcolor=blue}

\newcommand{\Tr}{\operatorname{Tr}}
\newcommand{\eff}{\mathrm{eff}}
\newcommand{\cN}{\mathcal{N}}

\begin{document}

\title{Clustered Attractor Manifolds and Dynamical Condensation in Self-Attention}

\author{Qucheng Gao}
\affiliation{Department of Physics, Boston College, Chestnut Hill, Massachusetts 02467, USA}
\author{Zuyi Yang}
\affiliation{Department of Physics, Tsinghua University,
Beijing 100084, China}
\author{Xiao Chen}
\affiliation{Department of Physics, Boston College, Chestnut Hill, Massachusetts 02467, USA}


\begin{abstract}
Transformer layers generate state-dependent interaction networks:
token representations determine the attention matrix, which in turn
updates the representations. We study this feedback in a minimal
normalized self-attention dynamics and identify the overlap gap as the
central quantity governing its attractor structure in the
thermodynamic limit. When tokens form internally aligned clusters and
their similarity to members of the same cluster exceeds that to every
other cluster by a nonvanishing amount, inter-cluster attention is
exponentially suppressed as the dimension increases. This mechanism
produces a high-dimensional manifold of clustered fixed points,
ranging from a few macroscopic clusters to extensive microscopic
fragmentation, and also controls their stability against
perturbations. Starting from an unstructured Gaussian state, we find
that clustered states nucleate from the diffuse background only above
a finite threshold in attention sharpness, giving rise to a dynamical
attention-condensation transition.
\end{abstract}

\maketitle

\emph{Introduction.--}
Transformers have become a central architecture in language, vision,
multimodal learning, and scientific inference
~\cite{vaswani2017attention,devlin2019bert,brown2020language,
dosovitskiy2021image,radford2021learning,jumper2021highly}.
Their defining operation, self-attention, converts pairwise token
similarities into row-wise softmax distributions. From a
statistical-mechanical perspective, each row is a Boltzmann
distribution with inverse temperature $\beta$. Moreover, transformer
depth defines a natural discrete-time dynamics
~\cite{lu2019multiparticle,bai2019deep},
\begin{equation}
X(t)\longrightarrow A[X(t)]\longrightarrow X(t+1),
\label{eq:feedback}
\end{equation}
in which the token configuration generates the interaction network,
which then feeds back on the tokens. This raises basic questions about
the attractors, stability, and phase structure of self-attention
dynamics.

Repeated attention layers can produce token uniformity and rank
collapse~\cite{dong2021rankcollapse,noci2022signal}, while strongly
localized attention has been linked to entropy collapse, training
instability, and attention sinks
~\cite{zhai2023entropy,xiao2024sinks}. Related mathematical work has
established clustering, mean-field limits, and nontrivial stationary
or multistable states in self-attention dynamics
~\cite{geshkovski2023clusters,geshkovski2025mathematical,
burger2025meanfield,Rigollet2026MeanField,altafini2025multistability,karagodin2024causal, karagodin2025normalization,bruno2025multiscale},
and random attention logits have been connected to the random-energy
model~\cite{giorlandino2026failure,derrida1981rem}. Here we show that,
in the joint limit $d\sim N\to\infty$, a minimal normalized
self-attention dynamics supports a high-dimensional manifold of
clustered fixed points controlled by the \emph{overlap gap}. A finite
gap generates an $O(\sqrt{N})$ logit advantage and exponentially
suppresses inter-cluster attention at fixed $\beta>0$. The resulting
manifold ranges from finitely many macroscopic clusters with diffuse
attention to extensive microscopic fragmentation with condensed
attention, while states with comparable attention localization can
have parametrically different representation geometries.

These clustered states are locally attracting: internal deformations
are damped, whereas collective cluster rotations move the system along
the fixed-point manifold. Finite perturbations may reorganize the
overlap gaps and transfer the dynamics between different attractors,
including fragmentation of a macroscopic cluster into microscopic
descendants. At finite $N$, residual inter-cluster attention produces
slow coarsening, but the lifetime of fragmented states grows
exponentially with $\beta\sqrt{N}$.

Finally, we ask which parts of the clustered attractor manifold are
dynamically accessible from an initially unstructured Gaussian cloud. At small $\beta$, broad attention
averaging erases the initial diversity and drives the representations
toward rank collapse, consistent with the known tendency of repeated
self-attention to produce token uniformity
~\cite{dong2021rankcollapse,noci2022signal}. Above a finite onset,
however, overlap fluctuations are amplified by the
attention--representation feedback, leading to the nucleation of
microscopic clusters and condensed attention.

\emph{Model and diagnostics.--}
We consider $N$ token vectors $x_i(t)\in\mathbb{R}^d$ with fixed norm
$\|x_i(t)\|=\sqrt d$.
Their normalized overlaps and scaled dot-product logits are
\begin{equation}
q_{ij}(t)=\frac{x_i(t)\cdot x_j(t)}{d},
\qquad
z_{ij}(t)=\sqrt d\,q_{ij}(t),
\qquad i\ne j.
\label{eq:overlap_logit}
\end{equation}
We mask the self-edge and define
\begin{equation}
A_{ij}(t)=
\frac{\exp[\beta z_{ij}(t)]}
{\sum_{k\ne i}\exp[\beta z_{ik}(t)]},
\qquad
A_{ii}=0,
\label{eq:attention}
\end{equation}
where $\beta$ controls attention sharpness. The residual normalized dynamics is
\begin{equation}
x_i(t+1)=
\cN\!\left[
(1-\gamma)x_i(t)+\gamma\sum_{j\ne i}A_{ij}(t)x_j(t)
\right],
\label{eq:dynamics}
\end{equation}
with $\cN[y]=\sqrt d\,y/\|y\|$. Unless noted otherwise, we take $d=N$ and
$\gamma=0.3$.

The row and mean attention inverse participation ratios (IPRs) are
\begin{equation}
Y_i(t)=\sum_{j\ne i}A_{ij}^2(t),
\qquad
Y_A(t)=\frac1N\sum_iY_i(t).
\label{eq:YA}
\end{equation}
Diffuse attention has $Y_A=O(1/N)$, whereas routing onto $O(1)$ targets gives
$Y_A=O(1)$. We characterize representation geometry through the normalized Gram matrix
$Q_{ij}=q_{ij}$ and its participation rank
\begin{equation}
R_{\eff}(t)=\frac{[\Tr Q(t)]^2}{\Tr[Q^2(t)]}
=\frac{N^2}{\sum_{ij}q_{ij}^2(t)}.
\label{eq:Reff}
\end{equation}

\emph{Clustered attractor manifold.--}
We first characterize the clustered fixed points of
Eq.~\eqref{eq:dynamics}. Partition the tokens into clusters
$\{\mathcal{C}_a\}_{a=1}^{K}$, with
$|\mathcal{C}_a|=n_a\geq 2$, and consider
\begin{equation}
x_i=\sqrt{d}\,c_a,
\qquad
i\in\mathcal{C}_a,
\qquad
Q_{ab}=c_a\cdot c_b,
\label{eq:cluster_state}
\end{equation}
where $c_a\cdot c_a=1$. The condition $n_a\geq 2$ follows from the
exclusion of diagonal self-attention.

For a source token in cluster $a$, the total attention assigned to
cluster $b$ is
\begin{equation}
P_{ab}
=
\frac{
(n_b-\delta_{ab})
\exp(\beta\sqrt{d}\,Q_{ab})
}{
\displaystyle
\sum_c
(n_c-\delta_{ac})
\exp(\beta\sqrt{d}\,Q_{ac})
}.
\label{eq:cluster_attention}
\end{equation}
Here $P_{aa}$ is the attention retained within the source cluster,
whereas $P_{ab}$ for $b\neq a$ is the inter-cluster leakage. Let 
\begin{equation} \Delta_a \equiv 1-\max_{b\neq a}Q_{ab} \label{eq:cluster_gap} \end{equation} 
be the overlap advantage of cluster $a$ over its closest competitor. The leakage vanishes if
\begin{equation}
\beta\sqrt{d}\,\Delta_a
-
\log\left(
\frac{N-n_a}{n_a-1}
\right)
\longrightarrow +\infty.
\label{eq:gap_fixed_point_condition}
\end{equation}
When this condition holds for every cluster, $P_{aa}\to 1$ and
$P_{ab}\to 0$ for $b\neq a$, so the configuration is a fixed point of
the limiting dynamics. For $d=N\to\infty$, any overlap gap bounded
away from zero satisfies this condition at fixed $\beta>0$. 

In the vanishing-leakage limit, a token in cluster $a$ attends
uniformly to its $n_a-1$ partners, giving
\begin{equation}
Y_A^{\star}
=
\frac{1}{N}
\sum_a
\frac{n_a}{n_a-1}.
\label{eq:cluster_ipr}
\end{equation}
Thus a finite number of macroscopic clusters has diffuse attention,
$Y_A^{\star}=O(N^{-1})$, whereas fixed-size microscopic clusters have
condensed attention, $Y_A^{\star}=O(1)$.

The representation geometry is characterized independently by
\begin{equation}
R_{\mathrm{eff}}^{\star}
=
\frac{N^2}{
\displaystyle
\sum_{a,b}
n_a n_b Q_{ab}^2
}.
\label{eq:cluster_reff}
\end{equation}
Broadly distributed microscopic clusters can have
$R_{\mathrm{eff}}=O(N)$, whereas clusters confined to a narrow cone
have $R_{\mathrm{eff}}=O(1)$. Thus similarly condensed attention
patterns can have parametrically different representation geometries. In the 
limit $N=d\to\infty$, these clustered fixed points are locally stable against
perturbations. 
Further details are given in the Supplemental Material~\cite{supp}.

\emph{Noise-induced fragmentation of a macroscopic cluster.--} 
Although clustered fixed points are locally stable against infinitesimal perturbations, a finite random perturbation can drive a macroscopic cluster beyond its basin of attraction and trigger fragmentation. Consider a single macroscopic parent cluster, 
\begin{equation} 
u_i(0) = \sqrt{1-\epsilon^2}\,u + \epsilon\,\eta_i, \qquad u\cdot\eta_i=0, \qquad \|\eta_i\|=1, \label{eq:noisy_consensus} 
\end{equation} 
where the $\eta_i$ are independent random directions in the tangent space at $u$. The pairwise overlaps are 
\begin{equation} q_{ij} = 1-\epsilon^2 + \epsilon^2\eta_i\cdot\eta_j. \label{eq:noisy_cluster_overlap} \end{equation} 
The common term $1-\epsilon^2$ cancels from each softmax row, leaving 
\begin{equation} A_{ij} = \frac{ \exp\!\left[ \alpha\sqrt d\,\eta_i\cdot\eta_j \right] }{ \displaystyle \sum_{k\neq i} \exp\!\left[ \alpha\sqrt d\,\eta_i\cdot\eta_k \right] }, \qquad \alpha\equiv\beta\epsilon^2. \label{eq:tangent_attention} \end{equation}
Thus the competition between collective averaging and selective routing is governed by the effective sharpness $\alpha=\beta\epsilon^2$. 

For small $\alpha$, attention remains nearly uniform. The random
tangent perturbations therefore cancel under averaging, and the tokens
return to a single aligned state. For large $\alpha$, even small
differences in pairwise angular overlap produce strong variations in
the attention weights. Each token then attends preferentially to nearby
directions, reinforcing local correlations and causing the parent
cluster to fragment into microscopic descendants.

Because these descendants originate from the same parent direction,
their centers remain confined to a common angular cone. The resulting
state has condensed attention while remaining geometrically close to
the rank-collapsed configuration. Once formed, the fragments are
protected by the overlap-gap condition
\begin{equation} \beta\sqrt N\,\Delta_N^{\rm cone} - \log N \longrightarrow+\infty, \label{eq:narrow_cone_stability} \end{equation} 
where $\Delta_N^{\rm cone}$ is the minimum same-cluster overlap advantage of the narrow-cone fragments.

\begin{figure}[t]
    \centering
    \includegraphics[width=\columnwidth]
    {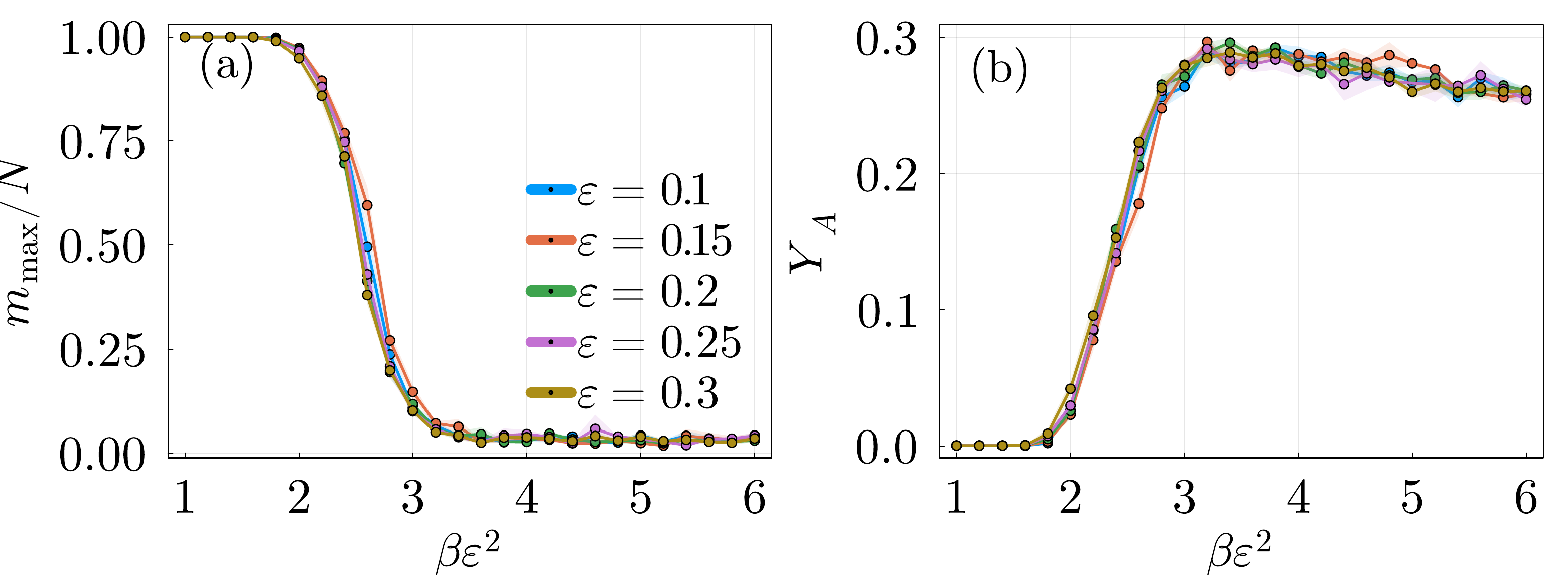}
    \caption{
    Noise-induced fragmentation of a macroscopic cluster for $N=4096$
    at observation time $T=N$.
    (a) Largest-cluster fraction $m_{\max}/N$.
    (b) Mean attention IPR $Y_A$.
    Results for five noise amplitudes are plotted against the effective sharpness $\alpha=\beta\epsilon^2$ and collapse onto a common curve. Shaded regions indicate the standard error of the mean across random realizations.
    }
    \label{fig:noisy_cluster_transition}
\end{figure}

Figure~\ref{fig:noisy_cluster_transition} shows this behavior for $N=4096$ and $T=N$. When plotted against $\alpha=\beta\epsilon^2$, results for five noise amplitudes collapse onto a common curve. We denote by $m_{\max}$ the size of the largest cluster; the clustering criterion is specified in the Supplemental Material~\cite{supp}. 
Below the onset, the largest cluster contains nearly all tokens,
\begin{equation}
\frac{m_{\max}}{N}\simeq1,
\qquad
Y_A=O(N^{-1}).
\end{equation}
Across the fragmentation region, $m_{\max}/N$ rapidly decreases while $Y_A$ becomes finite. At large $\alpha$, the largest cluster contains only a vanishing fraction of the tokens, whereas $Y_A=O(1)$, indicating an extensive collection of microscopic attention-condensed groups. The collapse identifies $\alpha$ as the relevant control parameter and supports a noise-induced fragmentation transition of the macroscopic cluster.

\emph{Stability of extensive fragmentation.--}
We now consider an extensive fragmented state containing
$K=\Theta(N)$ microscopic clusters of bounded size,
$n_a=O(1)$. Although each cluster has $O(N)$ possible external
targets, a nonvanishing overlap gap suppresses their total attention
exponentially in $\sqrt N$. Consequently, extensive fragmentation is
locally stable against weak random perturbations at every fixed
$\beta>0$ when $N\to\infty$ is taken first.

Consider first broadly distributed independent cluster centers in
$d=N$. Concentration of measure gives, with high probability,
\begin{equation}
\max_{a\neq b}Q_{ab}
=
O\!\left(\sqrt{\frac{\log N}{N}}\right),
\label{eq:broad_random_overlap}
\end{equation}
so distinct centers become asymptotically orthogonal and
\begin{equation}
\min_a\Delta_a
=
1-\max_{a\neq b}Q_{ab}
\longrightarrow1.
\label{eq:broad_random_gap}
\end{equation}
The overlap gap therefore remains finite and the inter-cluster
attention vanishes at every fixed $\beta>0$.

The same argument applies to narrow-cone fragmentation. Parameterize
the cluster centers as
\begin{equation}
c_a
=
\sqrt{1-w^2}\,u+w\eta_a,
\qquad
u\cdot\eta_a=0,
\qquad
\|\eta_a\|=1,
\label{eq:narrow_cone_centers}
\end{equation}
where $u$ is the common parent direction, $w$ is the cone width, and
the $\eta_a$ are independent random tangent directions. Their mutual
overlaps are
\begin{equation}
Q_{ab}
=
1-w^2+w^2\eta_a\cdot\eta_b.
\label{eq:narrow_cone_overlap}
\end{equation}
For $K=\Theta(N)$ random tangent directions,
$\max_{a\neq b}\eta_a\cdot\eta_b\to0$ as $N\to\infty$. Hence, for
any fixed $w>0$,
\begin{equation}
\min_a\Delta_a
\longrightarrow w^2
\label{eq:narrow_cone_gap}
\end{equation}
with high probability. Narrow-cone fragmentation therefore retains a
positive overlap gap, although it is generally smaller than that of
broad random fragmentation.

In both cases, the leakage bound derived above gives
\begin{equation}
\ell_a
\lesssim
N e^{-\beta\sqrt N\,\Delta_a}
\longrightarrow0
\label{eq:extensive_fragment_leakage}
\end{equation}
for every fixed $\beta>0$. The exponential suppression from the
overlap gap dominates the $O(N)$ number of competing clusters.
Therefore, each cluster dynamically decouples from the others in the
thermodynamic limit. Thus both broad random
fragmentation and narrow-cone fragmentation are locally stable against
weak perturbations at any fixed $\beta>0$ when the thermodynamic limit
is taken before the long-time limit.

\begin{figure}[t]
    \centering
    \includegraphics[width=\columnwidth]
    {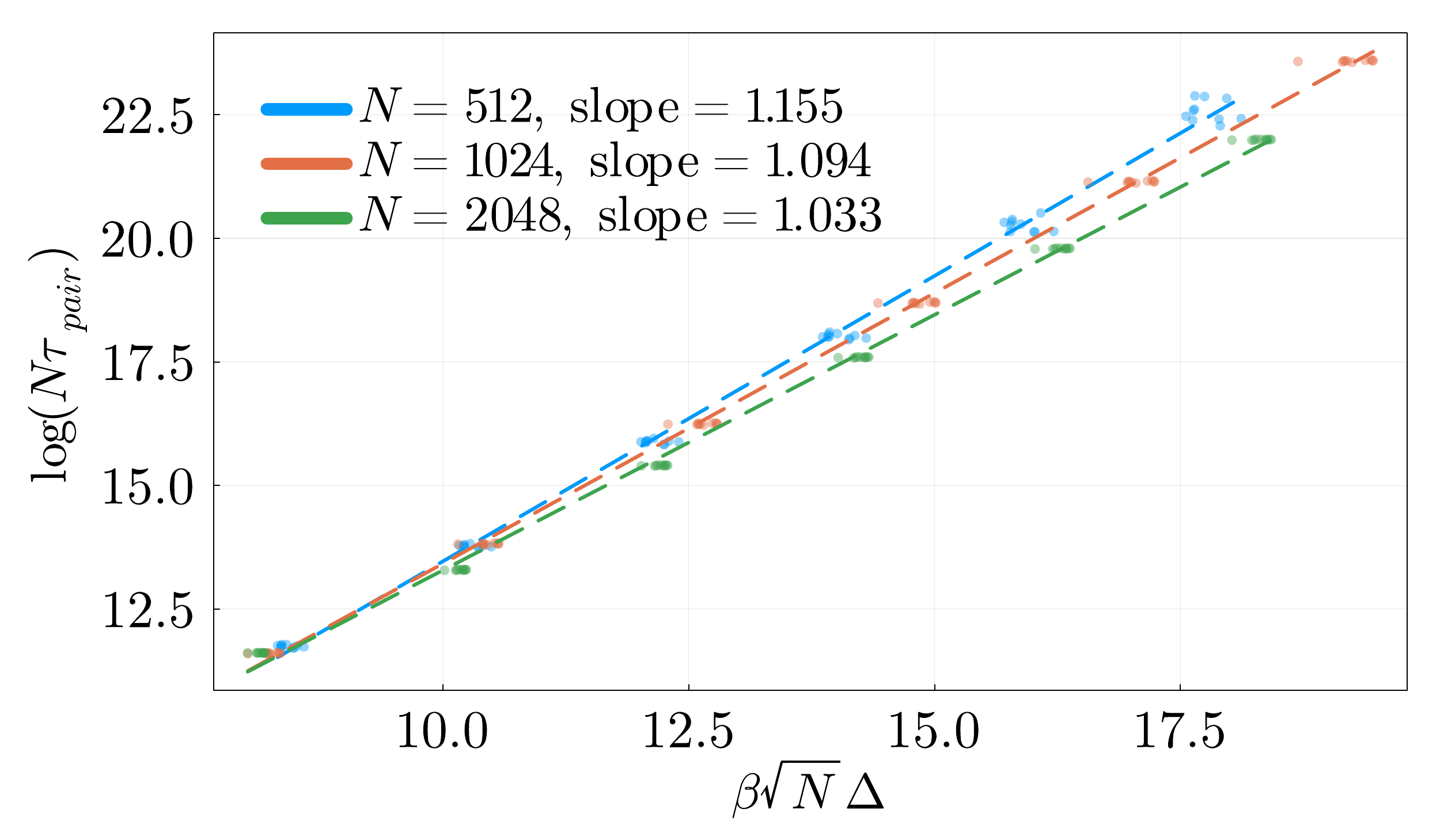}
    \caption{
    Scaling of the coarsening time for random pair clusters.
    The $N$ tokens are initialized as $N/2$ aligned pairs, with
    independently distributed random pair directions.
    Here $q_{pr}(t)$ denotes the overlap between pair-center directions
    $p$ and $r$, and
    $\Delta=1-\max_{p<r}q_{pr}(0)$ is the initial minimum pair gap.
    We define $\tau_{\rm pair}$ as the first time at which
    $\min_{p<r}q_{pr}(t)\geq q_{\rm con}$, with
    $q_{\rm con}=1-10^{-4}$.
    Results for $N=512$, $1024$, and $2048$ collapse when
    $\log(N\tau_{\rm pair})$ is plotted against
    $\beta\sqrt{N}\,\Delta$, consistent with the predicted unit slope.
    }
    \label{fig:pair_lifetime_scaling}
\end{figure}

At finite $N$, inter-cluster attention is exponentially small but
nonzero, producing a slow drift of the cluster centers. Microscopic
clusters can therefore merge into larger ones and may ultimately
approach consensus at sufficiently long times. For bounded-size
clusters, let $\Delta=\min_a\Delta_a$ denote the minimum overlap gap.
The fragmentation lifetime, set by the inverse leakage rate, scales as
\begin{equation} 
\tau_{\rm frag} \sim \frac{1}{\gamma N} \exp\!\left(\beta\sqrt N\,\Delta\right). \label{eq:fragment_lifetime_gap} \end{equation}
Finite systems can therefore coarsen at small $\beta$, but the
coarsening time grows exponentially with
$\beta\sqrt N\Delta$. This is confirmed in Figure~\ref{fig:pair_lifetime_scaling}. At any fixed $\beta>0$, this lifetime diverges
as $N\to\infty$. Consequently, the apparent finite-time stability
boundary shifts toward $\beta=0$ with increasing $N$, rather than
converging to a nonzero critical value.

The order of limits is therefore essential:
\begin{equation}
\lim_{N\to\infty}\lim_{t\to\infty}
\quad\hbox{versus}\quad
\lim_{t\to\infty}\lim_{N\to\infty}.
\label{eq:limits}
\end{equation}
Taking $t\to\infty$ first allows any finite system to coarsen, whereas
taking $N\to\infty$ first suppresses inter-cluster leakage and leaves
the extensively fragmented state dynamically stable.

\emph{Dynamical accessibility from Gaussian initial conditions.--}
We finally ask which parts of the clustered fixed-point manifold are
dynamically accessible from a generic unstructured state. We initialize
the tokens as independent normalized Gaussian vectors. For $d=N$,
\begin{equation}
q_{ij}(0)=O(N^{-1/2}),
\qquad
z_{ij}(0)=\sqrt N\,q_{ij}(0)=O(1).
\label{eq:gaussian_initial_scaling}
\end{equation}
Thus no token has a preexisting overlap advantage, while the softmax
acts on random $O(1)$ logits. After removing rowwise common offsets,
the initial competition is analogous to the noisy macroscopic-cluster
problem in Eq.~\eqref{eq:tangent_attention}, but without a common
parent direction and with effective sharpness $\beta$.

\begin{figure*}[t]
    \centering
    \includegraphics[width=\textwidth]{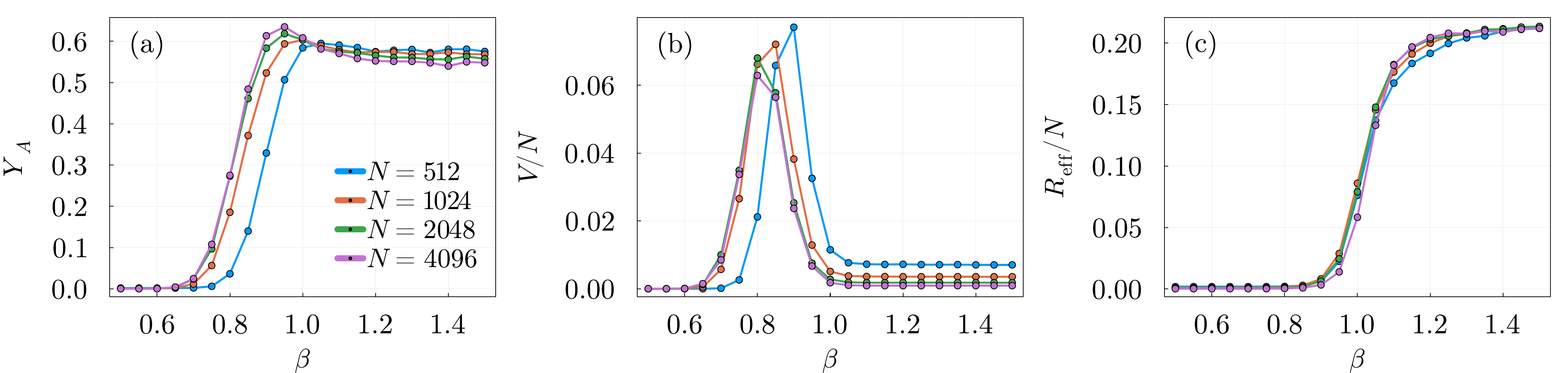}
    \caption{
    Dynamical accessibility from normalized Gaussian initial
    conditions with $T=N$.
    (a) The late-time attention IPR $Y_A$
    shows the onset of attention condensation.
    (b) The intermediate peak in $V/N$ signals a macroscopic spread of
    overlaps and the formation of a macroscopic cluster inside the condensation phase.
    (c) The effective rank $R_{\rm eff}/N$ becomes finite only at
    larger $\beta$, where the macroscopic cluster is replaced by
    extensive microscopic fragmentation.
    }
    \label{fig:caseA_numerics}
\end{figure*}

Figure~\ref{fig:caseA_numerics} shows the late-time observables as
$\beta$ is varied. At small $\beta$, attention remains diffuse and
broad averaging aligns the tokens,
\begin{equation}
Y_A=O(N^{-1}),
\qquad
\frac{R_{\eff}}{N}\to0.
\label{eq:diffuse_phase}
\end{equation}
The initially full-rank Gaussian cloud therefore flows toward a
rank-collapsed, nearly consensus state.

As $\beta$ increases, $Y_A$ becomes $O(1)$, signaling an
attention-condensation transition. Because clustered fixed points
already exist below this onset, the transition reflects the dynamical
nucleation of finite overlap gaps from the initially unstructured
cloud.

The condensed regime further separates into two geometrically distinct
states. To distinguish them, we consider the row-centered overlap
variance
\begin{equation}
\frac{V}{N}
=
\frac{1}{N(N-1)}
\sum_i\sum_{j\neq i}
\left(q_{ij}-\bar q_i\right)^2,
\quad
\bar q_i
=
\frac{1}{N-1}
\sum_{j\neq i}q_{ij}.
\label{eq:V_over_d_overlap}
\end{equation}
Immediately above the condensation onset, $V/N$ develops a pronounced
peak. Cluster diagnostics reveal one macroscopic high-overlap cluster
coexisting with an extensive number of microscopic condensed clusters.
Tokens in the macroscopic cluster distribute their attention over
$O(N)$ partners and have $Y_i=O(N^{-1})$, whereas the microscopic
clusters have
$Y_i=O(1)$ and generate the finite global IPR. The macroscopic cluster
also produces a dominant collective representation mode, keeping
$R_{\eff}$ subextensive. We refer to this coexistence state as the
\emph{macroscopic-clustered condensed regime}.

This macroscopic cluster emerges through size-biased coarsening. A
larger cluster receives more total attention because it contains more
possible targets. At intermediate $\beta$, inter-cluster leakage
remains strong enough to permit repeated mergers, allowing one early
cluster to become macroscopic, while sufficiently separated
microscopic clusters remain protected. Numerically, the formation time
of a tight macroscopic cluster grows as $O(\log N)$~\cite{Rigollet2026MeanField}; the same scaling
governs consensus formation in the diffuse regime. This logarithmic
time follows from the multiplicative amplification of the
$O(N^{-1/2})$ finite-size collective polarization and is described by
a logistic mean-field theory, as detailed in the Supplemental Material~\cite{supp}.

At larger $\beta$, attention remains condensed, but the macroscopic
cluster disappears:
\begin{equation}
Y_A=O(1),
\qquad
\frac{V}{N}\to0,
\qquad
\frac{R_{\eff}}{N}=O(1).
\label{eq:fragmented_phase}
\end{equation}
The representations instead form an extensive number of broadly
separated microscopic clusters. Each attention row remains localized
on $O(1)$ targets, but no representation mode carries macroscopic
weight. This is the \emph{fragmented condensed regime}. Here local
clusters form rapidly, while their overlap gaps suppress inter-cluster
leakage before macroscopic coarsening can occur.

The three observables therefore probe complementary aspects of the
dynamics: $Y_A$ detects attention condensation, $V/N$ detects
macroscopic overlap separation, and $R_{\eff}/N$ distinguishes a
dominant collective mode from extensive representation fragmentation.
As $\beta$ increases, the Gaussian initial state consequently accesses
three regimes: diffuse rank collapse, condensed coexistence with a
macroscopic cluster, and extensive microscopic fragmentation.

The condensation onset remains at finite $\beta$ as $N\to\infty$,
showing that it is a genuinely dynamical transition. For frozen
Gaussian logits with $O(1)$ variance, static softmax localization is
instead described by a random-energy-model benchmark and requires
$\beta_{\rm REM}\sim\sqrt{\log N}$~\cite{derrida1981rem,
giorlandino2026failure}. In the present dynamics, feedback between
attention and token geometry amplifies overlap fluctuations and
generates finite overlap gaps, enabling condensation at
$\beta=O(1)$. Details of the static benchmark are given in the
Supplemental Material~\cite{supp}.

\emph{Discussion.--}
We have shown that minimal self-attention supports a rich,
initial-condition-dependent attractor structure, ranging from diffuse rank
collapse to macroscopic condensation and extensive fragmentation. These
states are distinguished by their attention localization, cluster
statistics, and representation geometry.

Our analysis focuses on the scaling regime $d\sim N$. An $O(1)$ difference
in normalized overlap then produces a logit advantage
$\Delta z=O(\sqrt{N})$, allowing dynamically generated structure to
overcome the entropy of $O(N)$ competing targets at fixed $\beta$. If $d$
is held finite as $N\to\infty$, the logit advantage remains $O(1)$, and
fixed-$\beta$ condensation is not expected to survive in the same form.

The exponential selectivity of softmax is equally important. A logit
advantage $\Delta z$ produces a weight ratio
$\exp(\beta\Delta z)$, which can overcome the number of competing targets
when $d\sim N$. A linear attention kernel provides only algebraic
enhancement and therefore does not generate $O(1)$ attention IPR through
the same mechanism. Adaptive feedback alone is thus insufficient;
dynamical condensation also requires sufficiently nonlinear selection.
Finite-$d$ scaling and linear-attention dynamics are discussed further in
the Supplemental Material~\cite{supp}.

The minimal dynamics studied here provides a starting point for more
realistic transformer architectures. Nontrivial query, key, and value maps
should preserve the competition between averaging and nonlinear selection
while allowing additional condensed structures. Multi-head attention may
support coexistence or specialization among diffuse and condensed heads.
Applying the diagnostics developed here to such models may clarify how
architecture, depth, and learning shape the dynamical phases of attention.

\begin{acknowledgments}
The authors acknowledge the use of OpenAI's ChatGPT (GPT-5.5) for
brainstorming, drafting assistance, and exploratory analytical and numerical
calculations.  The authors independently verified the results and take full
responsibility for the manuscript.
\end{acknowledgments}

\nocite{
geshkovski2024metastability,
bruno2025metastable,
katharopoulos2020transformers,
choromanski2021rethinking}

\bibliography{trajectory_attention_v2_refs}

\end{document}


\title{Supplemental Material for ``Clustered Attractor Manifolds and Dynamical Condensation in Self-Attention''}

\author{Qucheng Gao}
\affiliation{Department of Physics, Boston College, Chestnut Hill, Massachusetts 02467, USA}
\author{Zuyi Yang}
\affiliation{Department of Physics, Tsinghua University,
Beijing 100084, China}
\author{Xiao Chen}
\affiliation{Department of Physics, Boston College, Chestnut Hill, Massachusetts 02467, USA}

\maketitle

\section{Static REM benchmark and IPR}
\label{secS:staticREM}

This section gives the density-of-states derivation behind the static REM benchmark~\cite{derrida1981rem} and explains how the row IPR detects the static transition.

\subsection{Density of states}

Consider one static row with fixed centered Gaussian logits
\begin{equation}
    u_j\sim\mathcal N(0,\sigma_{st}^2),
    \qquad j=1,\ldots,M.
\end{equation}
The row partition function is
\begin{equation}
    Z(\beta)=\sum_{j=1}^M e^{\beta u_j}.
\end{equation}
Let
\begin{equation}
    L=\log M,
    \qquad
    \epsilon=\frac{u}{\sigma_{st}\sqrt L}.
\end{equation}
The expected number of logits in $[\epsilon,\epsilon+d\epsilon]$ is
\begin{align}
    \Omega(\epsilon)d\epsilon
    &\simeq
    M\frac{d u}{\sqrt{2\pi}\sigma_{st}}
    \exp\left[-\frac{u^2}{2\sigma_{st}^2}\right]
    \\
    &=
    \sqrt{\frac{L}{2\pi}}
    \exp\left[L\left(1-\frac{\epsilon^2}{2}\right)\right]d\epsilon.
\end{align}
At exponential accuracy,
\begin{equation}
    \Omega(\epsilon)\asymp e^{L\Sigma(\epsilon)},
    \qquad
    \Sigma(\epsilon)=1-\frac{\epsilon^2}{2}.
    \label{eqS:Sigma}
\end{equation}
The typical upper edge of the density of states is
\begin{equation}
    \epsilon_{\rm edge}=\sqrt2,
    \qquad
    u_{\rm edge}=\sigma_{st}\sqrt{2\log M}.
\end{equation}

\subsection{Free-energy saddle}

For a nontrivial static REM competition set
\begin{equation}
    \beta=b\sqrt L.
\end{equation}
Grouping the softmax sum by scaled logit gives
\begin{equation}
    Z(\beta)\sim\int_{-\sqrt2}^{\sqrt2}d\epsilon\,
    \exp\left\{L\left[\Sigma(\epsilon)+b\sigma_{st}\epsilon\right]\right\}.
    \label{eqS:Zsaddle}
\end{equation}
The free-energy exponent is
\begin{equation}
    \phi(b)=\max_{\epsilon\le\sqrt2}\left[1-\frac{\epsilon^2}{2}+b\sigma_{st}\epsilon\right].
\end{equation}
The unconstrained saddle is $\epsilon_*=b\sigma_{st}$.  If $b\sigma_{st}<\sqrt2$, the saddle lies in the positive-entropy region and
\begin{equation}
    \phi(b)=1+\frac{b^2\sigma_{st}^2}{2}.
\end{equation}
The static REM transition occurs when $\epsilon_*=\sqrt2$, giving
\begin{equation}
    b_c=\frac{\sqrt2}{\sigma_{st}},
    \qquad
    \beta_{\REM}(M)=\frac{\sqrt{2\log M}}{\sigma_{st}}.
    \label{eqS:betaREM}
\end{equation}
For $b>b_c$, the saddle is pinned at the edge and
\begin{equation}
    \phi(b)=b\sigma_{st}\sqrt2.
\end{equation}

\subsection{IPR from free energies}

The row IPR is
\begin{equation}
    Y=\sum_j A_j^2=\frac{Z(2\beta)}{Z^2(\beta)}.
    \label{eqS:Ystatic}
\end{equation}
For $\beta=b\sqrt L$,
\begin{equation}
    \log Y\sim L\left[\phi(2b)-2\phi(b)\right].
\end{equation}
A direct evaluation gives
\begin{equation}
    \phi(2b)-2\phi(b)=
    \begin{cases}
    -1+b^2\sigma_{st}^2,
    &0\le b\le b_c/2,\\[3pt]
    -\left(b\sigma_{st}-\sqrt2\right)^2,
    &b_c/2\le b\le b_c,\\[3pt]
    0,
    &b\ge b_c.
    \end{cases}
    \label{eqS:psi_piecewise}
\end{equation}
Thus $Y\to0$ below the REM transition and $Y=O(1)$ in the frozen REM phase.  At fixed $\beta=O(1)$, the law of large numbers gives
\begin{equation}
    Z(\beta)\simeq M e^{\beta^2\sigma_{st}^2/2},
    \qquad
    Z(2\beta)\simeq M e^{2\beta^2\sigma_{st}^2},
\end{equation}
so
\begin{equation}
    Y\simeq \frac{e^{\beta^2\sigma_{st}^2}}{M}\to0.
    \label{eqS:Ystatic_fixedbeta}
\end{equation}
This is the benchmark: fixed finite $\beta$ initial random attention is diffuse.

\section{Condensation from dynamically generated clusters}
\label{secS:dynamic_cluster}

\subsection{Softmax condensation criterion}

Consider a query token $i$ for which the dynamics has generated
$k=O(1)$ preferred targets. Let their typical normalized overlap exceed that
of the remaining $M-k$ background targets by
\begin{equation}
    \Delta q_i
    =
    q_i^{\rm sel}-q_i^{\rm bg}.
\end{equation}
Since $z_{ij}=\sqrt d\,q_{ij}$, the corresponding logit gap is
\begin{equation}
    \Delta z_i
    =
    \sqrt d\,\Delta q_i.
    \label{eqS:dynamic_logit_gap}
\end{equation}
Softmax is invariant under a uniform shift of all logits in a row, so we may
set the background logit to zero. In a two-level approximation, the row then
contains $M-k$ background weights equal to unity and $k$ preferred weights
equal to $e^{\beta\Delta z_i}$. The partition function and row IPR are
\begin{equation}
    Z_i
    \simeq
    (M-k)+k e^{\beta\Delta z_i},
\end{equation}
and
\begin{equation}
    Y_i
    \simeq
    \frac{
        (M-k)+k e^{2\beta\Delta z_i}
    }{
        \left[
            (M-k)+k e^{\beta\Delta z_i}
        \right]^2
    }.
    \label{eqS:Ycluster_softmax}
\end{equation}
When
\begin{equation}
    k e^{\beta\Delta z_i}\ll M-k,
\end{equation}
the background dominates and $Y_i\simeq1/M$. The row condenses when the
preferred targets dominate,
\begin{equation}
    k e^{\beta\Delta z_i}\gtrsim M-k,
\end{equation}
or equivalently
\begin{equation}
    \beta\Delta z_i
    \gtrsim
    \log\left(\frac{M-k}{k}\right)
    \simeq
    \log\left(\frac{M}{k}\right).
    \label{eqS:cluster_condition}
\end{equation}
If the preferred targets have comparable logits, the condensed row has
$Y_i\simeq1/k=O(1)$.

For a finite dynamically generated overlap gap,
$\Delta q_i=O(1)$, Eq.~\eqref{eqS:dynamic_logit_gap} gives
$\Delta z_i=O(\sqrt d)$. In the scaling regime $d\sim M\sim N$, this logit
advantage grows as $\sqrt N$ and therefore exceeds the entropic scale
$\log N$ at any fixed $\beta>0$ for sufficiently large $N$. This argument
is conditional on the formation of a finite-overlap cluster; the role of the
dynamics is precisely to generate and stabilize such clusters. Moreover,
$Y_A=O(1)$ requires this row-level condensation to occur for an extensive
number of queries.

By contrast, the fixed-$d$ limit does not support condensation at fixed
$\beta$. Since $q_{ij}\in[-1,1]$, the logits have a bounded range,
$z_{ij}\in[-\sqrt d,\sqrt d]$. For a row containing $M$ available
targets,
\begin{equation}
Y_i
\leq
\max_{j\neq i} A_{ij}
\leq
\frac{\exp(2\beta\sqrt d)}{M}.
\end{equation}
Thus, for fixed $d$ and $\beta$, the row IPR vanishes as
$M\to\infty$, irrespective of the nonlinear feedback generated by the
dynamics. Finite-$d$ systems may still develop long-lived geometric
clusters at finite $N$, but obtaining an $O(1)$ row IPR requires the
sharpness to increase with system size. Consistently, the two-level
criterion gives
$\beta\sqrt d\Delta q_i\gtrsim\log(M/k)$, so the apparent
condensation crossover shifts to progressively larger $\beta$ as
$N$ increases rather than approaching a finite critical value.

\subsection{Clustering without condensation at fixed $d$.}
The absence of attention condensation does not preclude geometric
clustering. At fixed $d$ and fixed $\beta$, each individual attention
weight is $O(N^{-1})$. However, a macroscopic cluster
$\mathcal C_b$ containing $n_b=\rho_bN$ tokens can receive an
$O(1)$ total attention weight. Indeed, for a configuration with
$K=O(1)$ cluster centers $c_a$,
\begin{equation}
P_{ab}
=
\frac{
\rho_b\exp\left(\beta\sqrt d\,Q_{ab}\right)
}{
\displaystyle
\sum_c
\rho_c\exp\left(\beta\sqrt d\,Q_{ac}\right)
},
\qquad
Q_{ab}=c_a\cdot c_b.
\end{equation}
Similarity-dependent averaging can therefore synchronize an extensive
number of tokens into a finite number of macroscopic angular clusters,
even though the attention remains diffuse over the individual tokens.
For approximately uniform attention within each cluster, this gives
$Y_A\simeq K/N\to0$.

An exactly clustered configuration remains on the clustered manifold,
because all tokens within the same cluster receive identical updates.
Its centers evolve according to
\begin{equation}
c_a'
=
\frac{
(1-\gamma)c_a+\gamma\sum_bP_{ab}c_b
}{
\left\|
(1-\gamma)c_a+\gamma\sum_bP_{ab}c_b
\right\|
}.
\end{equation}
A generic clustered configuration is therefore not a fixed point at
finite $d$: the nonzero inter-cluster weights drive slow motion and
coarsening of the centers. A fixed point requires
\begin{equation}
\left(I-c_ac_a^{\mathsf T}\right)
\sum_bP_{ab}c_b
=
0
\end{equation}
for every cluster. Consensus always satisfies this condition, while
other fixed points require special self-consistent center geometries.

At large $\beta\sqrt d$, the interaction between well-separated
clusters is exponentially weak, so multicluster configurations may
remain metastable for very long times. Related finite-dimensional
self-attention models were shown to exhibit clustering and
exponentially long-lived multicluster states in
Refs.~\cite{geshkovski2023clusters,geshkovski2024metastability};
see also Ref.~\cite{bruno2025metastable} for a mean-field analysis.
The crucial distinction from the $d\sim N$ limit is that these
finite-$d$ clusters are generally macroscopic and have
$Y_A\sim K/N$, whereas microscopic clusters cannot dynamically
decouple at fixed $\beta$ as $N\to\infty$.

\subsection{Linear and polynomial attention kernels}

To isolate the role of nonlinear amplification, we first compare softmax with
a simple affine kernel motivated by linear-attention constructions
~\cite{katharopoulos2020transformers,choromanski2021rethinking}. For the same
two-level row, take
\begin{equation}
    K_{\rm lin}(u)=1+\beta u,
\end{equation}
where $u=0$ for the background targets and
$u=\Delta z_i>0$ for the $k=O(1)$ preferred targets. The normalized row has
\begin{equation}
    Z_i^{\rm lin}
    =
    (M-k)+k(1+\beta\Delta z_i)
    =
    M+k\beta\Delta z_i,
\end{equation}
and
\begin{equation}
    Y_i^{\rm lin}
    =
    \frac{
        (M-k)+k(1+\beta\Delta z_i)^2
    }{
        \left[M+k\beta\Delta z_i\right]^2
    }.
    \label{eqS:Ycluster_linear}
\end{equation}
At fixed $\beta$, with $d\sim M\sim N$, $k=O(1)$, and
$\Delta q_i=O(1)$, one has $\Delta z_i=O(\sqrt N)$. Both the numerator and
the correction to the denominator are then at most $O(N)$, while the squared
denominator is $O(N^2)$. Consequently,
\begin{equation}
    Y_i^{\rm lin}=O(N^{-1}),
\end{equation}
and affine attention does not condense at fixed $\beta$.

More generally, consider a positive polynomial kernel
\begin{equation}
    K_p(u)
    =
    1+[\beta u]_+^p,
    \qquad
    [u]_+=\max(u,0),
\end{equation}
with $p>0$. Defining
\begin{equation}
    g_i=[\beta\Delta z_i]^p,
\end{equation}
the two-level row gives
\begin{equation}
    Z_i^{(p)}
    =
    M+k g_i
\end{equation}
and
\begin{equation}
    Y_i^{(p)}
    =
    \frac{
        (M-k)+k(1+g_i)^2
    }{
        \left(M+k g_i\right)^2
    }.
    \label{eqS:Ycluster_power}
\end{equation}
For a finite overlap gap and $d\sim N$,
\begin{equation}
    \Delta z_i=O(\sqrt N),
    \qquad
    g_i=O(N^{p/2}).
\end{equation}
It follows that
\begin{equation}
    Y_i^{(p)}
    \sim
    \begin{cases}
        N^{-1}, & 0<p\leq1,\\[2pt]
        N^{p-2}, & 1<p<2,\\[2pt]
        O(1), & p=2,\\[2pt]
        1/k, & p>2.
    \end{cases}
    \label{eqS:power_kernel_scaling}
\end{equation}
Thus the quadratic kernel is the threshold case for condensation in the
scaling regime $d\sim N$. For $p=2$, the preferred targets carry a finite
fraction of the row weight and $Y_i=O(1)$, whereas for $p>2$ they dominate
the row and $Y_i\to1/k$. The exponential form of softmax is therefore not
strictly necessary. What is required is nonlinear amplification strong
enough for the weight of a finite-overlap target to scale at least as
$O(N)$.

\section{Clustered states: fixed points and their stability analysis}
\label{sec:supp_cluster_stability}

In this section we distinguish three related questions. First, we derive
the linear stability of an arbitrary exact clustered fixed point. Second,
we study a finite number of macroscopic parent clusters subject to a small
but finite random width and explain the transition to narrow-cone
fragmentation at large attention sharpness. Third, we show that an
extensive fragmentation into bounded-size clusters is stable at every
fixed $\beta>0$ in the ordered thermodynamic limit $d=N\to\infty$, even
though the same state can coarsen at finite $N$.

Throughout this section, ``stable at any finite $\beta$'' means any fixed
$\beta>0$, with $N\to\infty$ taken before the long-time limit. The point
$\beta=0$ is excluded. We use $\epsilon$ for the amplitude of an imposed
random perturbation and reserve
\begin{equation}
\ell_a
\equiv
\frac{1-P_{aa}}{P_{aa}}
=
\sum_{b\neq a}
\frac{n_b}{n_a-1}
e^{-\beta\sqrt d(1-Q_{ab})}
\end{equation}
for the dimensionless inter-cluster leakage of an exact cluster state.

\subsection{Clustered states and inter-cluster leakage}
\label{sec:supp_cluster_fixed_points}

We first summarize the attention structure of an exactly aligned
clustered configuration. Partition the $N$ tokens into clusters
$\{\mathcal C_a\}_{a=1}^{K}$, with
$|\mathcal C_a|=n_a\geq2$, and write
\begin{equation}
x_i=\sqrt d\,c_a,
\qquad
i\in\mathcal C_a,
\qquad
\|c_a\|=1.
\label{eq:supp_exact_cluster_state}
\end{equation}
We denote the overlap between cluster centers by
\begin{equation}
Q_{ab}=c_a\cdot c_b,
\qquad
Q_{aa}=1.
\label{eq:supp_cluster_overlap}
\end{equation}

For a source token $i\in\mathcal C_a$, define
\begin{equation}
P_{ab}
\equiv
\sum_{\substack{j\in\mathcal C_b\\j\neq i}}A_{ij}.
\label{eq:supp_Pab_definition}
\end{equation}
Thus $P_{ab}$ is the total attention in one attention row, associated
with a token in cluster $a$, that is assigned collectively to cluster
$b$. It is not summed over all source tokens in cluster $a$.
Internal alignment makes $P_{ab}$ independent of the particular
choice of $i\in\mathcal C_a$.

In particular,
\begin{equation}
P_{aa}
=
\sum_{\substack{j\in\mathcal C_a\\j\neq i}}A_{ij}
\label{eq:supp_Paa_definition}
\end{equation}
is the fraction of the attention row retained inside the source
cluster. Because the diagonal edge is masked, $P_{aa}$ is distributed
among the other $n_a-1$ tokens and does not include $A_{ii}$.
For $b\neq a$, $P_{ab}$ is the total attention assigned to all $n_b$
tokens in cluster $b$. The coarse-grained attention matrix is row
stochastic,
\begin{equation}
\sum_bP_{ab}=1.
\label{eq:supp_Pab_normalization}
\end{equation}

Every allowed edge from a source token in cluster $a$ to a target
token in cluster $b$ has the same logit,
\begin{equation}
z_{ij}
=
\frac{x_i\cdot x_j}{\sqrt d}
=
\sqrt d\,Q_{ab}.
\end{equation}
There are $n_b-\delta_{ab}$ allowed targets in cluster $b$.
Consequently,
\begin{equation}
P_{ab}
=
\frac{
(n_b-\delta_{ab})
e^{\beta\sqrt d\,Q_{ab}}
}{
\displaystyle
\sum_c
(n_c-\delta_{ac})
e^{\beta\sqrt d\,Q_{ac}}
}.
\label{eq:supp_cluster_attention}
\end{equation}
The attention weight on a single edge is therefore
\begin{equation}
A_{ij}
=
\frac{P_{ab}}{n_b-\delta_{ab}},
\qquad
i\in\mathcal C_a,\quad
j\in\mathcal C_b.
\label{eq:supp_single_edge_attention}
\end{equation}
Although $Q_{ab}=Q_{ba}$, one generally has
$P_{ab}\neq P_{ba}$ because the target multiplicities $n_a$ and
$n_b$ can be different.

Since $Q_{aa}=1$, the ratio of the total attention assigned to a
distinct cluster $b$ and the attention retained inside cluster $a$ is
\begin{equation}
\frac{P_{ab}}{P_{aa}}
=
\frac{n_b}{n_a-1}
e^{-\beta\sqrt d(1-Q_{ab})},
\qquad
b\neq a.
\label{eq:supp_relative_leakage}
\end{equation}
This motivates the definition
\begin{equation}
\ell_a
\equiv
\frac{1-P_{aa}}{P_{aa}}
=
\sum_{b\neq a}
\frac{n_b}{n_a-1}
e^{-\beta\sqrt d(1-Q_{ab})}.
\label{eq:supp_leakage_parameter}
\end{equation}
We refer to $\ell_a$ as the \emph{relative inter-cluster leakage} from
cluster $a$. It compares the total attention sent outside cluster $a$
with the attention retained inside it.

The actual outgoing attention fraction is
\begin{equation}
p_a^{\rm out}
\equiv
1-P_{aa}
=
\sum_{b\neq a}P_{ab}.
\label{eq:supp_outgoing_attention_definition}
\end{equation}
Using Eq.~\eqref{eq:supp_leakage_parameter}, one obtains
\begin{equation}
P_{aa}
=
\frac{1}{1+\ell_a},
\qquad
p_a^{\rm out}
=
\frac{\ell_a}{1+\ell_a}.
\label{eq:supp_leakage_relations}
\end{equation}
Therefore, in the weak-leakage regime,
\begin{equation}
p_a^{\rm out}
=
\ell_a+O(\ell_a^2).
\label{eq:supp_weak_leakage_relation}
\end{equation}
The distinction between $\ell_a$ and $p_a^{\rm out}$ is useful:
$\ell_a$ has the simple exponential representation in
Eq.~\eqref{eq:supp_leakage_parameter}, while $p_a^{\rm out}$ is the
actual probability mass leaving the cluster.

The attention output seen by a token in cluster $a$ is
\begin{equation}
\sum_{j\neq i}A_{ij}x_j
=
\sqrt d\sum_bP_{ab}c_b.
\label{eq:supp_cluster_attention_output}
\end{equation}
The cluster centers therefore evolve according to
\begin{equation}
c_a'
=
\frac{
(1-\gamma)c_a+\gamma\sum_bP_{ab}c_b
}{
\left\|
(1-\gamma)c_a+\gamma\sum_bP_{ab}c_b
\right\|
}.
\label{eq:supp_cluster_center_map}
\end{equation}
When $\ell_a\to0$, one has
\begin{equation}
P_{aa}\to1,
\qquad
P_{ab}\to0
\quad (b\neq a),
\end{equation}
and Eq.~\eqref{eq:supp_cluster_center_map} reduces to
$c_a'=c_a$. The clustered configuration is then an exact fixed point
of the limiting dynamics.

For later use, let
\begin{equation}
\Delta_a
\equiv
\min_{b\neq a}(1-Q_{ab})
\label{eq:supp_minimum_gap_a}
\end{equation}
be the minimum angular gap from cluster $a$ to any other center.
Equation~\eqref{eq:supp_leakage_parameter} gives the bound
\begin{equation}
\ell_a
\leq
\frac{N-n_a}{n_a-1}
e^{-\beta\sqrt d\,\Delta_a}.
\label{eq:supp_leakage_bound_a}
\end{equation}
Thus, if
\begin{equation}
\beta\sqrt d\,\Delta_a
-
\log\frac{N-n_a}{n_a-1}
\longrightarrow+\infty,
\label{eq:supp_fixed_point_condition_a}
\end{equation}
then $\ell_a\to0$ and cluster $a$ decouples from all the other
clusters.

\subsubsection{Attention IPR and representation rank of clustered fixed points}
\label{sec:supp_cluster_observables}

We derive the attention IPR and effective representation rank quoted in
the main text. Consider an exact clustered fixed point with cluster
sizes $\{n_a\}$ and center overlaps
\begin{equation}
Q_{ab}=c_a\cdot c_b.
\end{equation}
In the vanishing-leakage limit, a token $i\in\mathcal C_a$ attends
uniformly to the other $n_a-1$ tokens in its own cluster:
\begin{equation}
A_{ij}
=
\begin{cases}
\dfrac{1}{n_a-1},
&
j\in\mathcal C_a,\quad j\neq i,
\\[2mm]
0,
&
j\notin\mathcal C_a.
\end{cases}
\label{eq:supp_fixed_cluster_attention}
\end{equation}
The row attention IPR is therefore
\begin{align}
Y_i
&=
\sum_{j\neq i}A_{ij}^2
\nonumber\\
&=
(n_a-1)
\left(\frac{1}{n_a-1}\right)^2
=
\frac{1}{n_a-1},
\qquad
i\in\mathcal C_a.
\label{eq:supp_row_cluster_ipr}
\end{align}
Averaging over all tokens gives
\begin{equation}
Y_A^\star
=
\frac{1}{N}\sum_iY_i
=
\frac{1}{N}
\sum_a\frac{n_a}{n_a-1}.
\label{eq:supp_cluster_ipr}
\end{equation}

This expression shows that attention condensation is controlled by the
cluster-size distribution. If there are $K=O(1)$ macroscopic clusters
with $n_a=\rho_aN$, then
\begin{equation}
Y_A^\star
=
\frac{K}{N}+O(N^{-2}),
\label{eq:supp_macro_cluster_ipr}
\end{equation}
and attention is diffuse. By contrast, for $K=N/m$ equal clusters of
fixed size $m$,
\begin{equation}
Y_A^\star
=
\frac{1}{m-1},
\label{eq:supp_micro_cluster_ipr}
\end{equation}
which remains finite as $N\to\infty$. Microscopic clustering therefore
corresponds to condensed attention.

We next derive the effective representation rank. Let
$\mathsf Q$ denote the token-level overlap matrix,
\begin{equation}
\mathsf Q_{ij}
=
\frac{x_i\cdot x_j}{d}.
\end{equation}
For $i\in\mathcal C_a$ and $j\in\mathcal C_b$,
\begin{equation}
\mathsf Q_{ij}=Q_{ab}.
\end{equation}
Since $\mathsf Q_{ii}=1$,
\begin{equation}
\operatorname{Tr}\mathsf Q=N.
\end{equation}
Moreover,
\begin{align}
\operatorname{Tr}\mathsf Q^2
&=
\sum_{i,j}\mathsf Q_{ij}^2
\nonumber\\
&=
\sum_{a,b}
n_an_bQ_{ab}^2.
\label{eq:supp_gram_second_moment}
\end{align}
The participation rank is consequently
\begin{equation}
R_{\eff}^\star
=
\frac{(\operatorname{Tr}\mathsf Q)^2}
{\operatorname{Tr}\mathsf Q^2}
=
\frac{N^2}
{\displaystyle\sum_{a,b}n_an_bQ_{ab}^2}.
\label{eq:supp_cluster_reff}
\end{equation}
Unlike $Y_A^\star$, which depends primarily on cluster sizes,
$R_{\eff}^\star$ depends on the geometry of the cluster centers.

For $K$ equal clusters of size $m=N/K$,
Eq.~\eqref{eq:supp_cluster_reff} becomes
\begin{align}
R_{\eff}^\star
&=
\frac{K^2}
{K+\displaystyle\sum_{a\neq b}Q_{ab}^2}
\nonumber\\
&=
\frac{K}
{1+(K-1)q_{2,\mathrm{off}}},
\label{eq:supp_equal_cluster_reff}
\end{align}
where
\begin{equation}
q_{2,\mathrm{off}}
=
\frac{1}{K(K-1)}
\sum_{a\neq b}Q_{ab}^2
\label{eq:supp_q2off}
\end{equation}
is the mean squared overlap between distinct cluster centers.

For independently distributed random centers in $d$ dimensions,
\begin{equation}
\left\langle Q_{ab}^2\right\rangle
=
\frac{1}{d},
\qquad a\neq b.
\end{equation}
Thus, for $d=N$ and $K=\Theta(N)$,
\begin{equation}
q_{2,\mathrm{off}}=O(N^{-1}),
\qquad
R_{\eff}^\star=\Theta(N).
\label{eq:supp_random_center_reff}
\end{equation}
An extensive collection of broadly separated microscopic clusters
therefore retains extensive representation diversity.

For comparison, consider microscopic centers confined to a common
angular cone,
\begin{equation}
c_a
=
\sqrt{r}\,u+\sqrt{1-r}\,\eta_a,
\qquad
u\cdot\eta_a=0,
\label{eq:supp_narrow_cone_param}
\end{equation}
where the $\eta_a$ are random unit vectors in the tangent space. For
distinct centers,
\begin{equation}
Q_{ab}
=
r+(1-r)\eta_a\cdot\eta_b
=
r+O(d^{-1/2}).
\end{equation}
Hence
\begin{equation}
q_{2,\mathrm{off}}
\longrightarrow r^2,
\end{equation}
and Eq.~\eqref{eq:supp_equal_cluster_reff} gives
\begin{equation}
R_{\eff}^\star
\longrightarrow
\frac{1}{r^2}
=
O(1)
\label{eq:supp_narrow_cone_reff}
\end{equation}
as $K\to\infty$. The ordinary algebraic rank may still be extensive,
but the common direction $u$ carries a macroscopic fraction of the
Gram-matrix weight, leaving the participation rank finite.

Equations~\eqref{eq:supp_cluster_ipr} and
\eqref{eq:supp_cluster_reff} therefore characterize different aspects
of a clustered fixed point. The cluster sizes determine how many
targets receive appreciable attention, while the center overlaps
determine how broadly the token representations explore feature space.
For example, broad and narrow-cone microscopic fragmentation both have
$Y_A^\star=O(1)$, but respectively have
$R_{\eff}=O(N)$ and $R_{\eff}=O(1)$.

\subsubsection{Finite N analysis}

At finite $N$, the outgoing attention is small but nonzero. Writing
\begin{equation}
p_a^{\rm out}=1-P_{aa},
\end{equation}
the vector before normalization is
\begin{equation}
\widetilde c_a'
=
c_a+
\gamma\sum_{b\neq a}P_{ab}(c_b-c_a).
\end{equation}
For a unit vector $c_a$, normalization projects the perturbation onto
the tangent space at $c_a$. To leading order,
\begin{equation}
c_a'-c_a
=
\gamma\sum_{b\neq a}P_{ab}
\left(c_b-Q_{ab}c_a\right)
+
O\!\left[(\gamma p_a^{\rm out})^2\right].
\label{eq:supp_cluster_center_drift}
\end{equation}
The vector $c_b-Q_{ab}c_a$ is tangent to the sphere and has magnitude
$\sqrt{1-Q_{ab}^2}$. Define the geometric factor
\begin{equation}
g_a
=
\left\|
\sum_{b\neq a}
\frac{P_{ab}}{p_a^{\rm out}}
\left(c_b-Q_{ab}c_a\right)
\right\|.
\label{eq:supp_geometric_drift_factor}
\end{equation}
The one-step angular displacement is then
\begin{equation}
\delta\theta_a
=
\gamma p_a^{\rm out}g_a
+
O\!\left[(\gamma p_a^{\rm out})^2\right].
\label{eq:supp_angular_drift}
\end{equation}
For a generic configuration in which one or a few leakage channels
dominate, $g_a=O(1)$, and the inverse-leakage time scale is
\begin{equation}
\tau_a
\sim
\frac{1}{\gamma p_a^{\rm out}g_a}
=
\frac{1+\ell_a}{\gamma\ell_ag_a}
\simeq
\frac{1}{\gamma\ell_ag_a}.
\label{eq:supp_cluster_lifetime}
\end{equation}
$\tau_a$ is the characteristic time over which weak external attention produces an appreciable change in the cluster-center configuration.

If the drift is dominated by one cluster $b$, then
$g_a\simeq\sqrt{1-Q_{ab}^2}$ and
\begin{equation}
\tau_{a\leftarrow b}
\sim
\frac{n_a-1}
{\gamma n_b\sqrt{1-Q_{ab}^2}}
e^{\beta\sqrt d(1-Q_{ab})}.
\label{eq:supp_pairwise_lifetime}
\end{equation}
For a finite angular gap, the geometric factor is $O(1)$, leaving the
dominant exponential dependence
$\tau_{a\leftarrow b}\propto
e^{\beta\sqrt d(1-Q_{ab})}$.

\subsection{Linear stability of an arbitrary clustered fixed point}
\label{sec:supp_linear_stability}

Let $u_i=x_i/\sqrt d$ denote the unit-normalized token vectors. For an
exact clustered state,
\begin{equation}
u_i=c_a,
\qquad
i\in\mathcal C_a,
\qquad
\|c_a\|=1.
\label{eq:supp_exact_cluster}
\end{equation}
We perturb the tokens tangentially to the sphere,
\begin{equation}
u_i
=
c_a+\delta_i-\frac12\|\delta_i\|^2c_a+O(\delta^3),
\qquad
c_a\cdot\delta_i=0.
\label{eq:supp_normalized_perturbation}
\end{equation}
The radial term in Eq.~\eqref{eq:supp_normalized_perturbation} enforces
$\|u_i\|=1$ to second order. Decompose
\begin{equation}
\bar\delta_a
=
\frac1{n_a}\sum_{i\in\mathcal C_a}\delta_i,
\qquad
\xi_i
=
\delta_i-\bar\delta_a,
\qquad
\sum_{i\in\mathcal C_a}\xi_i=0.
\label{eq:supp_mode_decomposition}
\end{equation}
The collective mode $\bar\delta_a$ rotates the entire cluster, whereas
$\xi_i$ measures the internal spreading around the cluster center.

For two tokens in the same cluster,
\begin{align}
u_i\cdot u_j
&=
1-\frac12\|\delta_i\|^2-\frac12\|\delta_j\|^2
+\delta_i\cdot\delta_j+O(\delta^3)
\nonumber\\
&=
1-\frac12\|\delta_i-\delta_j\|^2+O(\delta^3).
\label{eq:supp_overlap_second_order}
\end{align}
There is no correction linear in $\delta$: a tangent displacement
changes the mutual overlap only at second order. Consequently, provided
that the perturbation is inside the local basin and inter-cluster leakage
is small,
\begin{equation}
A_{ij}
=
\frac{1}{n_a-1}
+
O\!\left(\beta\sqrt d\,\delta^2,\ell_a\right),
\qquad
i,j\in\mathcal C_a,\quad i\neq j.
\label{eq:supp_uniform_intra_attention}
\end{equation}
For random isotropic perturbations the typical centered logit variation
is smaller than the worst-case estimate in
Eq.~\eqref{eq:supp_uniform_intra_attention}; this point is discussed below.

The uniform average of the other $n_a-1$ members of cluster $a$ is
\begin{align}
\frac1{n_a-1}
\sum_{\substack{j\in\mathcal C_a\\j\neq i}}u_j
&=
c_a+
\frac{n_a\bar\delta_a-\delta_i}{n_a-1}
+O(\delta^2)
\nonumber\\
&=
c_a+\bar\delta_a-\frac{\xi_i}{n_a-1}
+O(\delta^2).
\label{eq:supp_average_other_tokens}
\end{align}
The minus sign has a simple origin: since the zero-mean internal modes
satisfy $\sum_i\xi_i=0$, omitting token $i$ from the average leaves
$\sum_{j\neq i}\xi_j=-\xi_i$.

Before the final normalization, the residual update is therefore
\begin{align}
\widetilde u_i'
&=
(1-\gamma)(c_a+\bar\delta_a+\xi_i)
+
\gamma\left(
c_a+\bar\delta_a-\frac{\xi_i}{n_a-1}
\right)
+
O(\delta^2,\ell_a\delta,\ell_a)
\nonumber\\
&=
c_a+\bar\delta_a+\lambda_a\xi_i
+
O(\delta^2,\ell_a\delta,\ell_a),
\label{eq:supp_pre_normalized_map}
\end{align}
where
\begin{equation}
\lambda_a
=
1-\gamma\frac{n_a}{n_a-1}.
\label{eq:supp_lambda}
\end{equation}
For a unit vector $c_a$ and a small displacement $h$,
\begin{equation}
\frac{c_a+h}{\|c_a+h\|}
=
c_a+\left(I-c_ac_a^{\mathsf T}\right)h+O(h^2).
\label{eq:supp_normalization_expansion}
\end{equation}
Since $\bar\delta_a$ and $\xi_i$ are tangent to $c_a$, normalization
does not alter them at linear order. Hence
\begin{align}
\bar\delta_a'
&=
\bar\delta_a+O(\delta^2,\ell_a),
\label{eq:supp_collective_mode}
\\
\xi_i'
&=
\lambda_a\xi_i
+
O(\delta^2,\ell_a\delta).
\label{eq:supp_internal_mode}
\end{align}

The collective mode is neutral when $\ell_a=0$: it simply moves the
configuration to a nearby point on the clustered fixed-point manifold.
The internal modes contract when
\begin{equation}
|\lambda_a|<1
\quad\Longleftrightarrow\quad
0<\gamma<\frac{2(n_a-1)}{n_a}.
\label{eq:supp_stability_condition}
\end{equation}
Thus $0<\gamma<1$ is sufficient to stabilize every cluster size
$n_a\geq2$. Defining the internal angular width
\begin{equation}
D_a
=
\frac1{n_a}
\sum_{i\in\mathcal C_a}\|\xi_i\|^2,
\label{eq:supp_cluster_width}
\end{equation}
one obtains
\begin{equation}
D_a'
=
\lambda_a^2D_a
+
O(D_a^{3/2},\ell_aD_a).
\label{eq:supp_width_decay}
\end{equation}
The clustered manifold is therefore normally attracting: perturbations
perpendicular to the manifold decay, while displacements along the
manifold are neutral in the thermodynamic limit.

\subsection{Finite random width of macroscopic parent clusters}
\label{sec:supp_macroscopic_noise}

We next consider $K_0=O(1)$ macroscopic parent clusters,
$n_a=\rho_aN$, whose center directions remain separated by finite
angles. Because inter-parent leakage is exponentially small in
$\sqrt d$, each parent cluster can be analyzed independently on the
time scale of its internal evolution.

For a single parent direction $u$, take
\begin{equation}
u_i(0)
=
\sqrt{1-\epsilon^2}\,u+\epsilon\,\eta_i,
\qquad
u\cdot\eta_i=0,
\qquad
\|\eta_i\|=1,
\label{eq:supp_noisy_macroscopic_cluster}
\end{equation}
where the $\eta_i$ are independent isotropic directions in the tangent
space. The pairwise overlaps are
\begin{equation}
q_{ij}
=
1-\epsilon^2+\epsilon^2\eta_i\cdot\eta_j.
\label{eq:supp_macroscopic_overlap}
\end{equation}
The row-independent term $1-\epsilon^2$ cancels from the softmax. The
attention within the parent cluster is therefore controlled by
\begin{equation}
A_{ij}^{(a)}
\simeq
\frac{
\exp\!\left[
\beta\sqrt d\,\epsilon^2\,\eta_i\cdot\eta_j
\right]
}{
\displaystyle
\sum_{\substack{k\in\mathcal C_a\\k\neq i}}
\exp\!\left[
\beta\sqrt d\,\epsilon^2\,\eta_i\cdot\eta_k
\right]
}.
\label{eq:supp_tangent_attention}
\end{equation}
For independent tangent directions,
\begin{equation}
\operatorname{Var}(\eta_i\cdot\eta_j)
=
\frac{1}{d-1}+o(d^{-1}),
\end{equation}
and hence the row-centered logit variance is
\begin{equation}
\sigma_z^2
\equiv
\operatorname{Var}_j
\!\left[
\sqrt d\,\epsilon^2\eta_i\cdot\eta_j
\right]
=
\epsilon^4[1+o(1)].
\label{eq:supp_logit_variance_noise}
\end{equation}
The typical routing inhomogeneity is therefore governed by
\begin{equation}
\alpha
\equiv
\beta\epsilon^2.
\label{eq:supp_effective_sharpness}
\end{equation}
The largest centered logit in a row acquires only an additional
$\sqrt{\log n_a}$ extreme-value factor.

When $\alpha\ll1$, attention is approximately uniform inside the parent
cluster. The random tangent directions average to zero and
Eq.~\eqref{eq:supp_width_decay} contracts the angular width. The parent
cluster returns to a single aligned direction, shifted only by the small
sample mean of the noise.

When $\alpha=O(1)$ or larger, token-dependent routing can become
heterogeneous before the angular width has relaxed. Tokens preferentially
select slightly more similar neighbors; the update increases their
mutual overlap and further sharpens the same attention edges. The
positive feedback can split a macroscopic parent cluster into many
microscopic descendants. Since the descendants originate from the same
parent direction, they remain inside its angular cone. For
$K_0=O(1)$ parent clusters, the nonlinear outcome is therefore a finite
number of narrow cones, each containing an extensive number of
microscopic routing groups.

This phenomenon is not a linear instability of the exact clustered
state. At every fixed finite $\beta$, the limit
$\epsilon\to0$ gives $\alpha\to0$, and the perturbation is healed.
For a fixed nonzero $\epsilon$, however, increasing $\beta$ can move the
state outside the local basin.

\subsection{Stability of extensive fragmentation in the presence of noise}
\label{sec:supp_extensive_fragmentation}

We finally consider $K=\Theta(N)$ microscopic clusters with bounded
sizes
\begin{equation}
2\le n_a\le m_{\max}=O(1),
\qquad
x_i=\sqrt N\,c_a,
\qquad
i\in\mathcal C_a,
\label{eq:supp_extensive_state}
\end{equation}
and independent random center directions $c_a$. With probability
tending to one,
\begin{equation}
q_{\max}(N)
\equiv
\max_{a\neq b}c_a\cdot c_b
=
O\!\left(\sqrt{\frac{\log N}{N}}\right)
=o(1).
\label{eq:supp_random_center_max}
\end{equation}
For an exact fragmented state, Eq.~\eqref{eq:supp_leakage_parameter}
therefore obeys
\begin{align}
\ell_a
&\leq
\frac{N}{n_a-1}
\exp\!\left[
-\beta\sqrt N(1-q_{\max})
\right]
\nonumber\\
&=
\exp\!\left[
-\beta\sqrt N
+O(\sqrt{\log N})
+\log N
\right]
\longrightarrow0
\label{eq:supp_extensive_leakage_bound}
\end{align}
for every fixed $\beta>0$. The extensive number of competing clusters
contributes only an entropy $\log N$, while the same-cluster logit
advantage is $O(\sqrt N)$. Thus the exact extensively fragmented state
is a fixed point of the limiting dynamics.

Infinitesimal random noise around each fragment is governed by the
linear analysis of Sec.~\ref{sec:supp_linear_stability}. The internal
width of every fragment contracts according to
Eq.~\eqref{eq:supp_width_decay}, while the inter-fragment center motion
vanishes with $\ell_a$. Hence the extensively fragmented manifold is
locally stable at every fixed $\beta>0$.

A stronger finite-noise statement can be formulated in terms of a
surviving intra-cluster overlap advantage. For a source token $i$ in
cluster $a$, define
\begin{equation}
q_i^{\rm in}
=
\max_{\substack{j\in\mathcal C_a\\j\neq i}}q_{ij},
\qquad
q_i^{\rm out}
=
\max_{j\notin\mathcal C_a}q_{ij},
\qquad
\Delta_i=q_i^{\rm in}-q_i^{\rm out}.
\label{eq:supp_noisy_gap}
\end{equation}
If the perturbation preserves a finite gap
$\Delta_i\geq\Delta_0>0$, then the total attention leaving the original
fragment satisfies
\begin{equation}
P_i^{\rm out}
\leq
N\exp\!\left[-\beta\sqrt N\,\Delta_0\right]
\longrightarrow0
\label{eq:supp_noisy_out_bound}
\end{equation}
at every fixed $\beta>0$. Therefore different parent fragments cannot
merge in the thermodynamic limit. The microscopic partition is exactly
restored when the perturbation is inside the linear basin. Outside that
basin, a bounded-size parent fragment may split into still smaller
descendants, but Eq.~\eqref{eq:supp_noisy_out_bound} prevents mixing
between the broadly separated parent directions. Extensive
fragmentation is consequently robust even when the precise microscopic
partition is not.

Random pair clusters provide the cleanest finite-noise example because
each token has a unique intra-pair target. Let
\begin{equation}
u_{a,s}(0)
=
\sqrt{1-\epsilon^2}\,c_a+\epsilon\,\eta_{a,s},
\qquad
s=1,2,
\label{eq:supp_noisy_pair_state}
\end{equation}
with independent random $c_a$ and weak independent tangent noise. Denote
the partner overlap by $q_{\rm p}$ and a typical inter-pair overlap by
$r$. The overlap advantage is
\begin{equation}
\Delta q_{\rm p}=q_{\rm p}-r.
\label{eq:supp_pair_gap}
\end{equation}
As long as $\Delta q_{\rm p}$ remains positive and $O(1)$, the attention
retained on the partner is
\begin{equation}
P_{\rm p}
\simeq
\frac{1}{
1+(N-2)
e^{-\beta\sqrt N\,\Delta q_{\rm p}}
}.
\label{eq:supp_pair_retention}
\end{equation}
For every fixed $\beta>0$, $P_{\rm p}\to1$. Once the partner is selected,
the difference of the two token perturbations contracts as
\begin{equation}
\delta_{a,1}'-\delta_{a,2}'
=
(1-2\gamma)
(\delta_{a,1}-\delta_{a,2})
+o(1).
\label{eq:supp_pair_healing}
\end{equation}
Thus finite noise heals inside each pair, while attention exchanged with
other pairs vanishes.

At finite $N$, $P_i^{\rm out}$ is small but nonzero, and the fragment
centers undergo slow coarsening. The characteristic drift time is
\begin{equation}
\tau_a
\sim
\frac{1}{\gamma(1-P_{aa})}
\simeq
\frac{1}{\gamma\ell_a}.
\label{eq:supp_fragment_lifetime_general}
\end{equation}
For bounded-size random clusters,
\begin{equation}
\tau_{\rm frag}
\sim
\frac{1}{\gamma N}
\exp\!\left[\beta\sqrt N(1-o(1))\right].
\label{eq:supp_fragment_lifetime}
\end{equation}
The lifetime diverges faster than any power of $N$ at every fixed
$\beta>0$, but it remains finite for every finite $N$. This explains
why sufficiently long simulations at small $\beta$ can merge the
fragments into a single cluster.

For an observation time $T$, the operational survival condition
$\tau_{\rm frag}\gg T$ gives
\begin{equation}
\beta\sqrt N
\gtrsim
\log N+\log T+O(1),
\label{eq:supp_operational_condition}
\end{equation}
or
\begin{equation}
\beta_{\rm surv}(N,T)
\sim
\frac{\log N+\log T}{\sqrt N}
\longrightarrow0.
\label{eq:supp_operational_boundary}
\end{equation}
The apparent small-$\beta$ stability boundary at finite size and finite
time therefore moves to zero in the thermodynamic limit.

The order of limits is essential. Schematically,
\begin{equation}
\lim_{t\to\infty}\lim_{N\to\infty}
\quad
\text{retains the extensive fragmented attractor,}
\label{eq:supp_order_limits_1}
\end{equation}
whereas
\begin{equation}
\lim_{N\to\infty}\lim_{t\to\infty}
\quad
\text{can select consensus after finite-$N$ coarsening.}
\label{eq:supp_order_limits_2}
\end{equation}

In summary, the clustered manifold is linearly stable because uniform
intra-cluster averaging contracts every zero-mean internal deformation.
Finite random width can nevertheless fragment a macroscopic parent
cluster when attention becomes sufficiently sharp, producing microscopic
descendants inside the same narrow cone. By contrast, an extensive
fragmentation with broadly separated parent directions is protected by
an $O(\sqrt N)$ intra-cluster logit advantage: weak noise heals locally,
inter-fragment mixing vanishes at every fixed $\beta>0$, and finite-size
coarsening is exponentially slow in $\beta\sqrt N$.

\section{Dynamical formation of a macroscopic cluster from a Gaussian state}

\subsection{Numerical results}
\label{sec:gaussian_cluster_nucleation}

We investigate the time required for a macroscopic coherent structure
to emerge from a Gaussian random initial state. Writing
$x_i(t)=\sqrt{d}\,c_i(t)$ with $\|c_i(t)\|=1$, we monitor the global
polarization
\begin{equation}
s(t)
=
\left\|
\frac{1}{N}\sum_{i=1}^{N}c_i(t)
\right\|^2.
\label{eq:supp_polarization}
\end{equation}
For independent random initial directions, $s(0)\sim N^{-1}$, whereas
$s(t)=O(1)$ indicates that a finite fraction of the tokens has
developed a common directional component.

The polarization does not distinguish a tightly localized cluster
from a broader aligned cloud. We therefore also measure
\begin{equation}
p_q(t)
=
\frac{1}{N}
\max_i
\sum_{j=1}^{N}
\Theta\!\left[c_i(t)\cdot c_j(t)-q\right],
\label{eq:supp_cluster_fraction}
\end{equation}
which gives the largest fraction of tokens lying within an angular
neighborhood defined by the overlap threshold $q$. For $q$ close to
unity, $p_q=O(N^{-1})$ in a diffuse state and becomes $O(1)$ when a
tightly aligned macroscopic cluster forms. 

We simulate the full self-masked attention dynamics with $d=N$,
$\gamma=0.3$
and consider three representative cases: uniform attention
at $\beta=0$, diffuse consensus at $\beta=0.4$, and macroscopic-cluster
formation at $\beta=0.8$.

\begin{figure}[t]
\centering
\includegraphics[width=0.75\linewidth]
{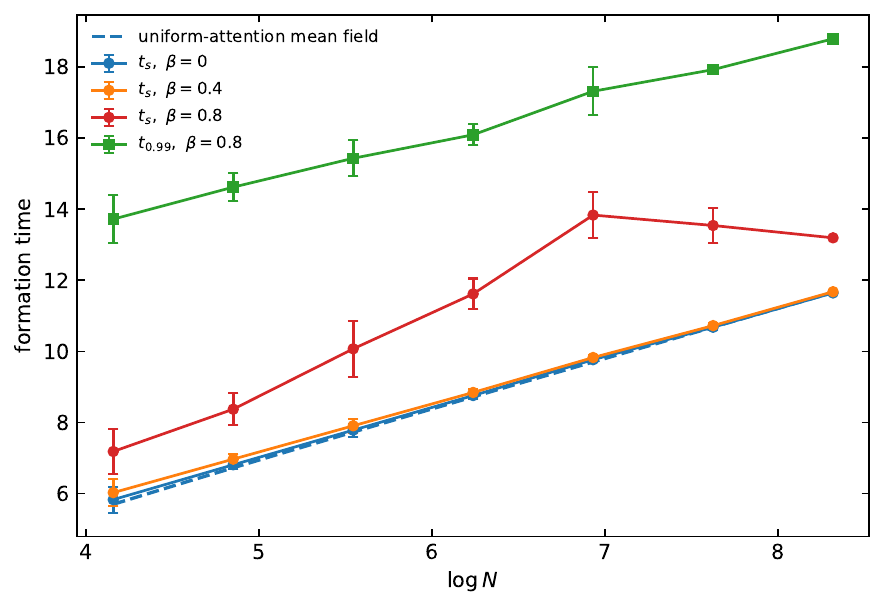}
\caption{Formation times from Gaussian random initial states. For
$\beta=0$ and $\beta=0.4$, we characterize the formation time by
$t_s=\inf\{t:s(t)\geq 1/2\}$, where
$s(t)=\left\|N^{-1}\sum_i c_i(t)\right\|^2$ is the global
polarization. Thus, $t_s$ marks the first time at which a macroscopic
collective component becomes $O(1)$. At $\beta=0.8$, we show two
different times. The polarization time $t_s$ again marks the onset of
broad collective alignment, while
$t_{0.99}=\inf\{t:p_{0.99}(t)\geq 1/2\}$ measures the later time at
which a tightly aligned macroscopic cluster forms. The separation between
$t_s$ and $t_{0.99}$ at $\beta=0.8$ shows that, in the intermediate
regime, the growth of a macroscopic coherent component and the final
sharpening of that component into a tight cluster are distinct
dynamical processes. The dashed line for the uniform-attention mean field theory is explained later in Sec.~\ref{sec:supp_log_time_origin}. Error bars denote the standard deviation over
random initial conditions.}
\label{fig:supp_nucleation_logN}
\end{figure}

\begin{figure}[t]
\centering
\includegraphics[width=0.75\linewidth]
{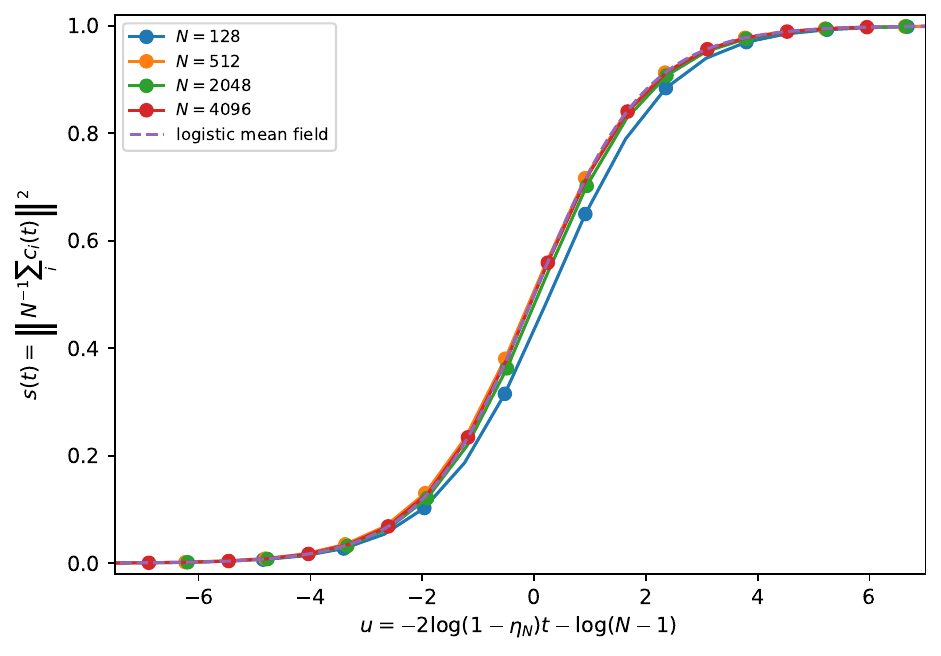}
\caption{Logistic collapse of the polarization dynamics at
$\beta=0.4$. Results for sizes up to $N=4096$ closely follow the
uniform-attention scaling function $s(u)=(1+e^{-u})^{-1}$.}
\label{fig:supp_logistic_collapse}
\end{figure}

The results are shown in Fig.~\ref{fig:supp_nucleation_logN}. For $\beta=0$ and $\beta=0.4$, the emerging structure is well
described by a single broad collective mode, so the polarization time
\begin{equation}
t_s=\inf\{t:s(t)\geq 1/2\}
\end{equation}
provides a natural definition of the formation time. Here
$s(t)$ measures the strength of
the global collective component. These two curves also collapse
when plotted against
\begin{equation}
u
=
-2\log(1-\eta_N)t-\log(N-1),
\label{eq:supp_logistic_scaling_variable}
\end{equation}
and closely follow the logistic scaling function
$s(u)=(1+e^{-u})^{-1}$, as shown in
Fig.~\ref{fig:supp_logistic_collapse}. This scaling function follows
from the uniform-attention mean-field theory developed in
Sec.~\ref{sec:supp_log_time_origin}.

\begin{figure}[t]
\centering
\includegraphics[width=0.75\linewidth]
{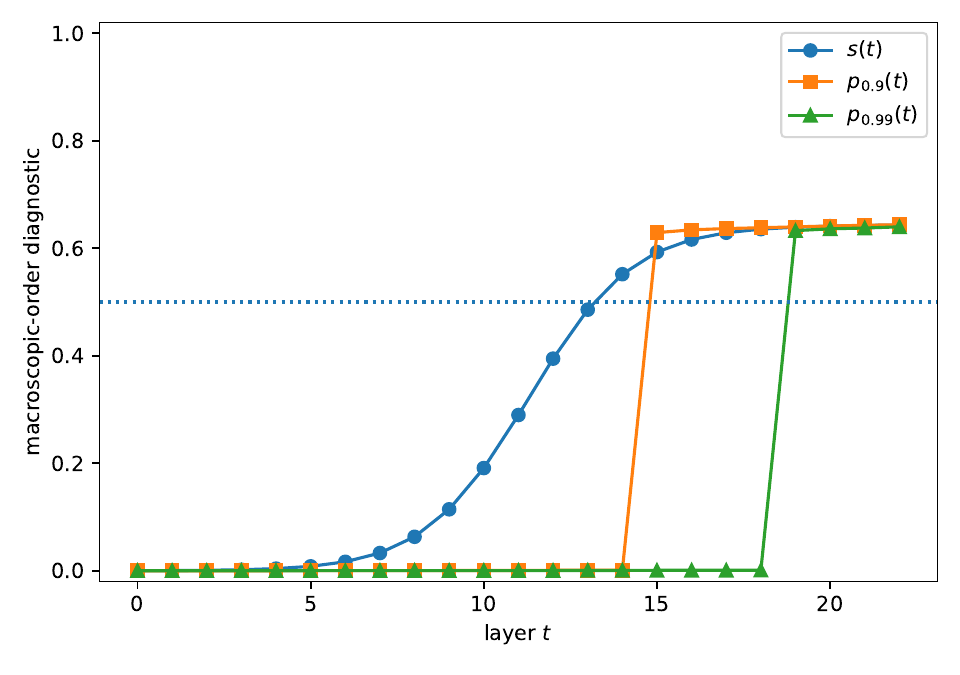}
\caption{Formation and sharpening of a macroscopic cluster for
$N=d=4096$ and $\beta=0.8$. The polarization $s(t)$ becomes
macroscopic before the high-overlap fractions $p_{0.9}(t)$ and
$p_{0.99}(t)$, separating collective alignment from the subsequent
formation and sharpening of a tight cluster.}
\label{fig:supp_cluster_growth_N4096}
\end{figure}

In the intermediate regime, $\beta=0.8$, the polarization becomes
$O(1)$ before the tokens condense into a geometrically tight
macroscopic cluster. We therefore show both $t_s$ and
\begin{equation}
t_{0.99}
=
\inf\{t:p_{0.99}(t)\geq 1/2\},
\end{equation}
where $p_{0.99}(t)$ denotes the largest fraction of tokens whose
overlap with a single reference-token direction exceeds $0.99$. The
time $t_s$ marks the emergence of a broad macroscopic coherent
component and exhibits visible deviations from a simple $\log N$
scaling. By contrast, $t_{0.99}$ characterizes the later sharpening of
this component into a tightly aligned cluster and shows a clearer
logarithmic dependence on $N$. The separation between these two times
demonstrates that collective alignment and cluster sharpening are
distinct dynamical processes in the intermediate regime. This
interpretation is further supported by the time-resolved results in
Fig.~\ref{fig:supp_cluster_growth_N4096}.

\subsection{Physical origin of the logarithmic formation time}
\label{sec:supp_log_time_origin}

The logarithmic formation time originates from the amplification of a
finite-size collective fluctuation. Define the mean token direction and
the corresponding polarization as
\begin{equation}
m(t)
=
\frac{1}{N}\sum_{i=1}^{N}c_i(t),
\qquad
s(t)=\|m(t)\|^2.
\label{eq:supp_mean_polarization}
\end{equation}
For independent random initial directions,
\begin{equation}
\langle s(0)\rangle=\frac{1}{N},
\qquad
\|m(0)\|\sim N^{-1/2}.
\label{eq:supp_initial_seed}
\end{equation}
Thus, although the Gaussian ensemble is isotropic on average, every
finite-size realization contains a small residual polarization along a
random direction. This fluctuation provides the seed for consensus or
for the growth of a dominant macroscopic cluster.

\paragraph{Uniform-attention mean field.--}
The amplification mechanism is particularly transparent at
$\beta=0$. In this case, the attention weights in the full transformer
dynamics are exactly uniform,
\begin{equation}
A_{ij}=\frac{1}{N-1},
\qquad i\neq j.
\label{eq:supp_uniform_attention}
\end{equation}
The unnormalized token update therefore becomes
\begin{align}
\widetilde c_i
&=
(1-\gamma)c_i
+
\frac{\gamma}{N-1}\sum_{j\neq i}c_j
\nonumber\\
&=
(1-\eta_N)c_i+\eta_N m,
\qquad
\eta_N=\frac{\gamma N}{N-1}.
\label{eq:supp_uniform_update}
\end{align}
Decomposing each token into its collective and fluctuating components,
\begin{equation}
c_i=m+\delta_i,
\qquad
\sum_i\delta_i=0,
\end{equation}
gives
\begin{equation}
\widetilde c_i
=
m+(1-\eta_N)\delta_i.
\label{eq:supp_mean_fluctuation_update}
\end{equation}
The uniform averaging step leaves the common component unchanged while
contracting the fluctuations around it by a factor $1-\eta_N$.
Normalizing each token back
to unit length then increases the relative weight of the common
component. Consequently,
\begin{equation}
\frac{\|m(t)\|}{\|\delta(t)\|}
\sim
N^{-1/2}(1-\eta_N)^{-t}.
\label{eq:supp_signal_noise_growth}
\end{equation}

To obtain a closed equation for the complete dynamics of $s(t)$, below we use the
permutation-symmetric mean-field approximation in which all tokens have
the same overlap with $m$ and therefore the same normalization factor~\cite{Rigollet2026MeanField}.
This closure is exact for an equiangular configuration and becomes
accurate for a self-averaging diffuse Gaussian cloud. This method is discussed in detail in Ref.~\cite{Rigollet2026MeanField}. The common
squared norm of the unnormalized tokens is
\begin{equation}
r_t^2
=
\|\widetilde c_i\|^2
=
(1-\eta_N)^2
+
\eta_N(2-\eta_N)s_t.
\label{eq:supp_uniform_norm}
\end{equation}
Since the unnormalized mean remains equal to $m(t)$, the normalization
step gives
\begin{equation}
s_{t+1}
=
\frac{s_t}{
(1-\eta_N)^2+\eta_N(2-\eta_N)s_t
}.
\label{eq:supp_discrete_polarization_map}
\end{equation}
Equivalently,
\begin{equation}
\frac{1-s_{t+1}}{s_{t+1}}
=
(1-\eta_N)^2
\frac{1-s_t}{s_t}.
\label{eq:supp_discrete_odds}
\end{equation}
Using $s_0=1/N$, this recursion has the solution
\begin{equation}
s_t
=
\frac{1}{
1+(N-1)(1-\eta_N)^{2t}
}.
\label{eq:supp_discrete_logistic}
\end{equation}

The dashed uniform-attention curve in
Fig.~\ref{fig:supp_nucleation_logN} is obtained analytically from
Eq.~\eqref{eq:supp_discrete_logistic}. The numerical points at $\beta=0$ are obtained by evolving
the full $N$-token dynamics. Although the attention weights are exactly
uniform in that simulation, finite-size Gaussian samples are not
perfectly permutation symmetric, so their normalization factors
fluctuate slightly from token to token. These fluctuations produce the
small deviations from the mean-field curve and vanish progressively as
$N$ increases.

Defining the formation time by $s(t_s)=1/2$, the discrete mean-field
solution gives
\begin{equation}
t_s
=
\frac{\log(N-1)}
{-2\log|1-\eta_N|}
=
\frac{\log N}
{-2\log(1-\gamma)}
+O(1).
\label{eq:supp_log_time_prediction}
\end{equation}
The logarithmic dependence has a simple interpretation. During the
early stage, $s_t\ll1$, Eq.~\eqref{eq:supp_discrete_logistic} reduces
to
\begin{equation}
s_t
\simeq
\frac{1}{N}(1-\eta_N)^{-2t}
=
\frac{1}{N}e^{\lambda_N t},
\qquad
\lambda_N=-2\log|1-\eta_N|.
\label{eq:supp_early_discrete_growth}
\end{equation}
The initial collective weight is $O(N^{-1})$, while every layer
amplifies it by an $O(1)$ multiplicative factor. Reaching
$s_t=O(1)$ therefore requires
\begin{equation}
e^{\lambda_N t_{\mathrm{form}}}\sim N,
\qquad
t_{\mathrm{form}}\sim\lambda_N^{-1}\log N.
\end{equation}

For a small residual step, $\gamma\ll1$, the discrete map reduces to
the continuous logistic equation
\begin{equation}
\frac{ds}{dt}
=
2\gamma s(1-s).
\label{eq:logistic_polarization}
\end{equation}
Its solution for $s(0)=1/N$ is
\begin{equation}
s(t)
=
\frac{1}{
1+(N-1)e^{-2\gamma t}
}.
\label{eq:logistic_solution}
\end{equation}
The factor $s$ represents the positive feedback generated by the
preexisting coherent component, whereas $1-s$ represents the
remaining incoherent weight available for alignment. The growth is
exponential while $s\ll1$ and slows only when the polarization becomes
macroscopic.

Weak attention heterogeneity can modify the growth rate without
changing this mechanism. In the intermediate regime, however, the
global polarization and the geometric compactness of the cluster
describe different stages of the dynamics. The time $t_s$ marks the
appearance of a broad macroscopic coherent component, while
$t_{0.99}$ measures the later sharpening of this component into a
tightly aligned cluster. The uniform-attention theory predicts the
former collective amplification process but not the additional
sharpening time.

Finally, the logarithmic macroscopic formation time should be
distinguished from two other dynamical scales. At larger $\beta$,
microscopic routing groups can emerge within $O(1)$ layers directly
from fluctuations of the initial attention weights. Conversely, the
subsequent coarsening or merging of well-separated clusters can occur
on a much longer finite-size time scale. The resulting hierarchy is
\begin{equation}
t_{\mathrm{micro}}=O(1),
\qquad
t_{\mathrm{macro}}=O(\log N),
\qquad
t_{\mathrm{coarsen}}\gg\log N.
\label{eq:time_scale_hierarchy}
\end{equation}

\section{Additional Numerical Results}

\subsection{Snapshots from different initial conditions}
To visualize the geometric difference between the fragmented states
reached from different initial conditions, we compare snapshots of the
dynamics starting from Gaussian random and nearly rank-collapsed
configurations.

For the Gaussian initial condition, the tokens are sampled as
independent normalized Gaussian vectors,
\begin{equation}
    x_i(0)
    =
    \sqrt d\,\frac{g_i}{\|g_i\|},
    \qquad
    g_i\sim\mathcal N(0,I_d).
\end{equation}
For the nearly rank-collapsed initial condition, the tokens are weakly
dispersed around a common direction,
\begin{equation}
    x_i(0)
    =
    \mathcal N_{\sqrt d}
    \left[
        \sqrt{1-\epsilon^2}\,u+\epsilon\,\eta_i
    \right],
\end{equation}
where $u$ is a unit vector, the $\eta_i$ are random unit vectors
orthogonal to $u$, and $\mathcal N_{\sqrt d}$ denotes normalization to
norm $\sqrt d$.

We choose parameters for which both initial conditions evolve into
attention-condensed states containing an extensive number of
microscopic clusters. Despite this common local cluster structure, their
global representation geometries are markedly different. Starting from
the Gaussian state, the cluster directions remain broadly distributed
in representation space. Starting from the nearly rank-collapsed state,
the clusters remain concentrated within a narrow cone around the common
initial direction.

\begin{figure*}[t]
    \centering
    \includegraphics[width=\textwidth]
    {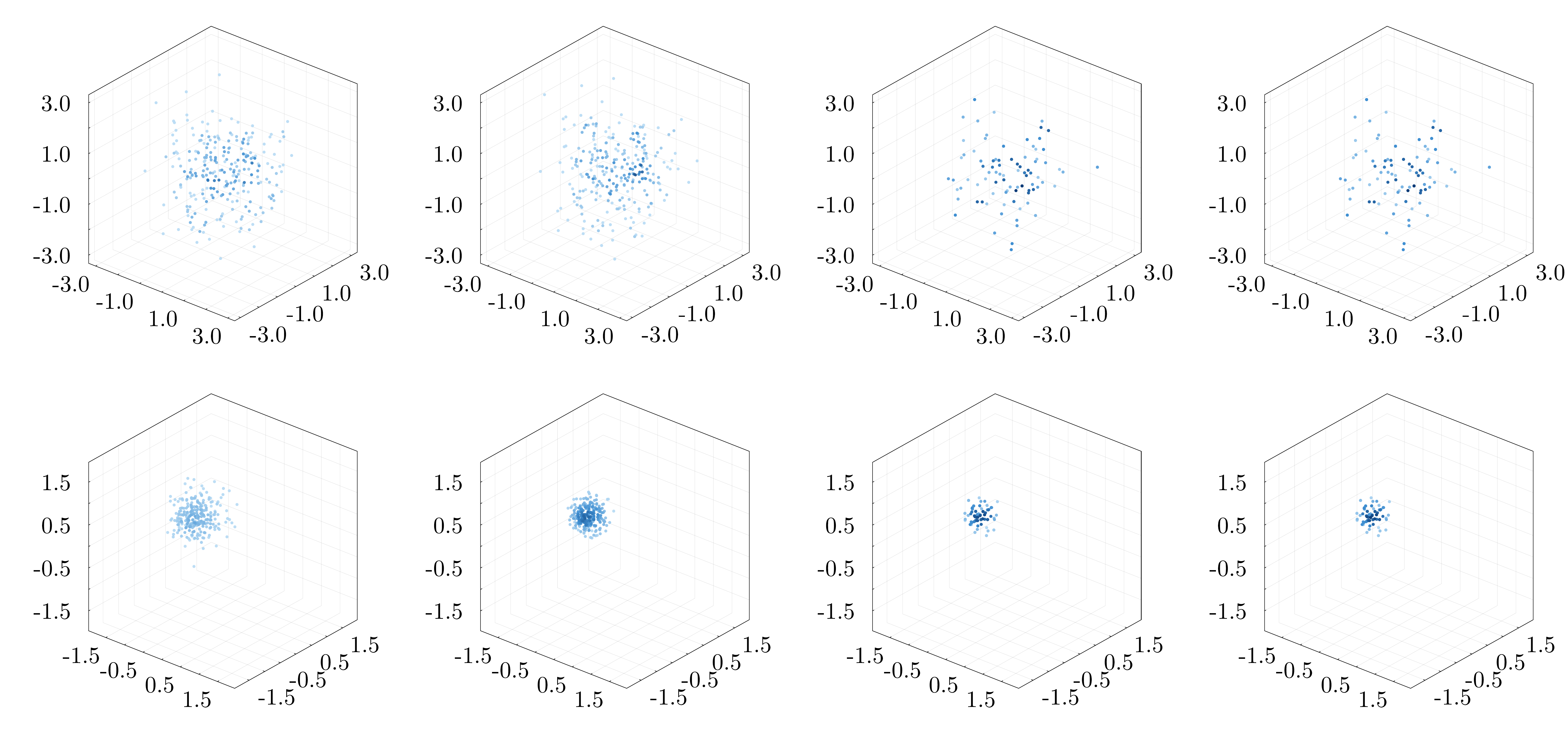}
    \caption{
    Snapshots of the dynamics under fixed three-dimensional random
    projections for $N=256$. From left to right, the snapshots are taken
    at $t=0$, $2$, $50$, and $300$. The upper row shows Gaussian random
    initialization with $\beta=1.5$. The lower row shows nearly
    rank-collapsed initialization with $\epsilon=0.3$ and
    $\beta\epsilon^2=6.0$. The same projection directions are used at
    all displayed times within each row. 
    Marker color indicates local token density in the projected three-dimensional space, with darker colors corresponding to denser regions.
    The two initial conditions
    produce, respectively, broad and narrow-cone fragmented geometries.
    }
    \label{fig:supp_random_projection}
\end{figure*}

Figure~\ref{fig:supp_random_projection} shows snapshots of the token
configurations under fixed three-dimensional random projections. The
same projection directions are used at all displayed times for each
initial condition, allowing the evolution of the global geometry to be
followed directly.

\subsection{Cluster statistics and threshold dependence}
To characterize the fragmented configurations, we measure the size of
the largest cluster, $m_{\max}$, and the total number of clusters,
$N_{\rm cl}$. Two tokens $i$ and $j$ are assigned to the same cluster
when their overlap satisfies
\begin{equation}
    q_{ij}>q_{\rm th}.
\end{equation}
Cluster membership is transitive: if $i$ and $j$ are assigned to the
same cluster and $j$ and $k$ are assigned to the same cluster, then
$i$, $j$, and $k$ are grouped together even when $q_{ik}$ does not
directly exceed the threshold.

\begin{figure*}[t]
    \centering
    \includegraphics[width=\textwidth]
    {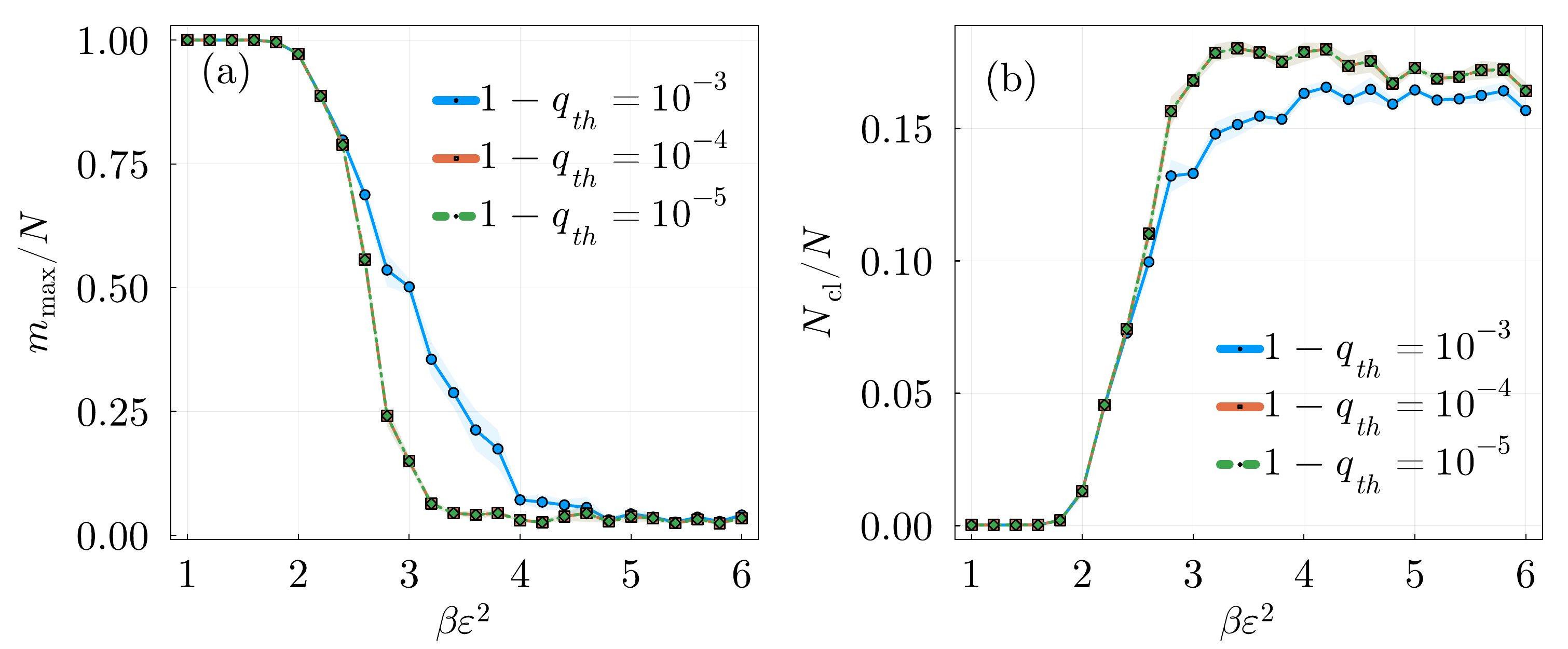}
    \caption{
    Robustness of the cluster statistics with respect to the overlap
    threshold $q_{\rm th}$ for nearly rank-collapsed initial conditions
    with $N=4096$ and $\epsilon=0.1$.
    (a) Largest cluster size $m_{\max}$.
    (b) Total number of clusters $N_{\rm cl}$.
    Results are shown for
    $1-q_{\rm th}=10^{-3}$, $10^{-4}$, and $10^{-5}$.
    Shaded regions indicate the standard error of the mean across
    random realizations.
    }
    \label{fig:supp_delta_comparison}
\end{figure*}

Because tokens within a dynamically formed cluster become nearly
aligned, the extracted cluster statistics should be insensitive to the
precise value of $q_{\rm th}$ when the threshold is sufficiently close
to unity. Figure~\ref{fig:supp_delta_comparison} tests this robustness
by comparing
\begin{equation}
    1-q_{\rm th}
    =
    10^{-3},\ 10^{-4},\ 10^{-5}.
\end{equation}
The results for $1-q_{\rm th}=10^{-4}$ and $10^{-5}$ are nearly
indistinguishable, whereas the coarser threshold
$1-q_{\rm th}=10^{-3}$ produces visible deviations by grouping some
nearby but distinct fragments together. We therefore use
\begin{equation}
    q_{\rm th}=1-10^{-4}
\end{equation}
for all cluster observables reported in the paper.

\section{Numerical characterization of the dynamical regimes}
\label{sec:supp_numerical_regimes}

We characterize the dynamics using six complementary observables that
probe attention localization, representation geometry, and cluster
structure. The mean attention inverse participation ratio is
\begin{equation}
    Y_A
    =
    \frac{1}{N}
    \sum_i\sum_{j\neq i}A_{ij}^2.
\end{equation}
Diffuse attention over $O(N)$ targets gives
$Y_A=O(N^{-1})$, whereas attention localized on $O(1)$ targets gives
$Y_A=O(1)$.

We also measure the mean row attention entropy,
\begin{equation}
    S_A
    =
    -\frac{1}{N}
    \sum_i\sum_{j\neq i}A_{ij}\log A_{ij},
\end{equation}
and report the normalized quantity $S_A/\log N$. For approximately
uniform attention, $S_A/\log N$ approaches unity as $N$ increases.
When each row is localized on a finite number of targets,
$S_A=O(1)$ and therefore $S_A/\log N\to0$. The observables $Y_A$ and
$S_A/\log N$ thus provide complementary measures of attention
condensation.

The global representation geometry is characterized by the
row-centered logit variance
\begin{equation}
    V
    =
    \frac{1}{N(N-1)}
    \sum_i\sum_{j\neq i}
    \left(z_{ij}-\bar z_i\right)^2,
    \qquad
    \bar z_i
    =
    \frac{1}{N-1}
    \sum_{j\neq i}z_{ij},
\end{equation}
and by the participation rank
\begin{equation}
    R_{\rm eff}
    =
    \frac{N^2}{\sum_{ij}q_{ij}^2}.
\end{equation}
Since $z_{ij}=\sqrt{N}\,q_{ij}$ for $d=N$, a finite value of
$V/N$ indicates an $O(1)$ spread of normalized overlaps across an
extensive fraction of a row. The ratio $R_{\rm eff}/N$ distinguishes
a representation dominated by a finite number of collective modes
from one with extensive geometric diversity.

Finally, the cluster structure is characterized by the largest cluster
size, $m_{\max}$, and the total number of clusters, $N_{\rm cl}$.
A finite value of $m_{\max}/N$ signals the presence of a macroscopic
cluster, whereas $N_{\rm cl}/N=O(1)$ indicates an extensive number of
microscopic clusters.

\subsection{Gaussian initial conditions}
For Gaussian initialization, we study
\begin{equation}
    Y_A,\qquad
    \frac{S_A}{\log N},\qquad
    \frac{V}{N},\qquad
    \frac{R_{\rm eff}}{N},\qquad
    \frac{m_{\max}}{N},\qquad
    \frac{N_{\rm cl}}{N}
\end{equation}
as functions of $\beta$ for
$N=512$, $1024$, $2048$, and $4096$. Unless stated otherwise, the
observables are measured at the common observation time $T=N$.

\begin{figure*}[t]
    \centering
    \includegraphics[width=\textwidth]
    {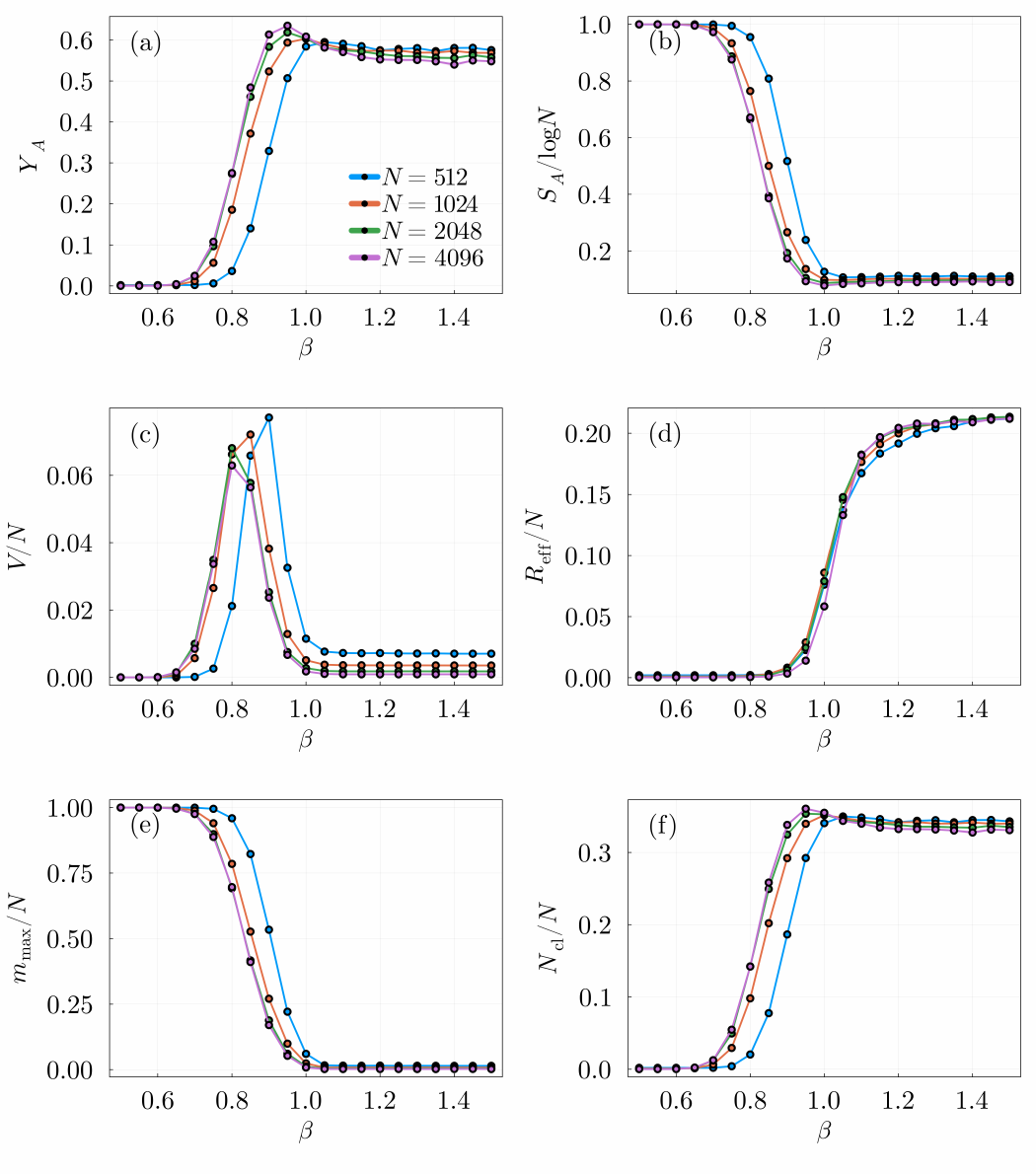}
    \caption{
    Attention, representation, and cluster observables for Gaussian
    initial conditions, measured at $T=N$ for
    $N=512$, $1024$, $2048$, and $4096$.
    The six panels show $Y_A$, $S_A/\log N$, $V/N$,
    $R_{\rm eff}/N$, $m_{\max}/N$, and $N_{\rm cl}/N$ as functions
    of $\beta$. Together, these observables distinguish the diffuse
    rank-collapsed, macroscopic-clustered condensed, and fragmented
    condensed regimes.
    }
    \label{fig:supp_gaussian_observables}
\end{figure*}

Figure~\ref{fig:supp_gaussian_observables} resolves the three dynamical
regimes identified in the main text. At small $\beta$, attention
remains diffuse: $Y_A$ decreases with increasing $N$, while
$S_A/\log N$ remains close to unity. The tokens align into a dominant
cluster, so $m_{\max}/N$ approaches unity, while
$R_{\rm eff}/N$ and $N_{\rm cl}/N$ vanish. This is the
diffuse rank-collapsed regime.

At intermediate $\beta$, attention becomes condensed:
$Y_A$ remains finite while $S_A/\log N$ decreases with increasing
system size. At the same time, both $m_{\max}/N$ and
$N_{\rm cl}/N$ remain finite. These observations show that one
macroscopic cluster coexists with an extensive collection of
microscopic clusters. The peak in $V/N$ reflects the macroscopic
separation between the large cluster and the remaining token
population. Because the macroscopic cluster produces a dominant
collective representation mode, $R_{\rm eff}/N$ remains small. This
is the macroscopic-clustered condensed regime.

At larger $\beta$, the macroscopic cluster disappears:
$m_{\max}/N$ decreases toward zero, while $N_{\rm cl}/N$ remains
finite. Attention remains condensed, as indicated by finite $Y_A$ and
small $S_A/\log N$. In this regime, $V/N$ decreases because no single
cluster contributes an extensive set of mutually large overlaps. By
contrast, $R_{\rm eff}/N$ becomes finite, showing that the microscopic
cluster centers are broadly distributed in representation space. This
is the fragmented condensed regime.

\begin{figure*}[t]
    \centering
    \includegraphics[width=\textwidth]
    {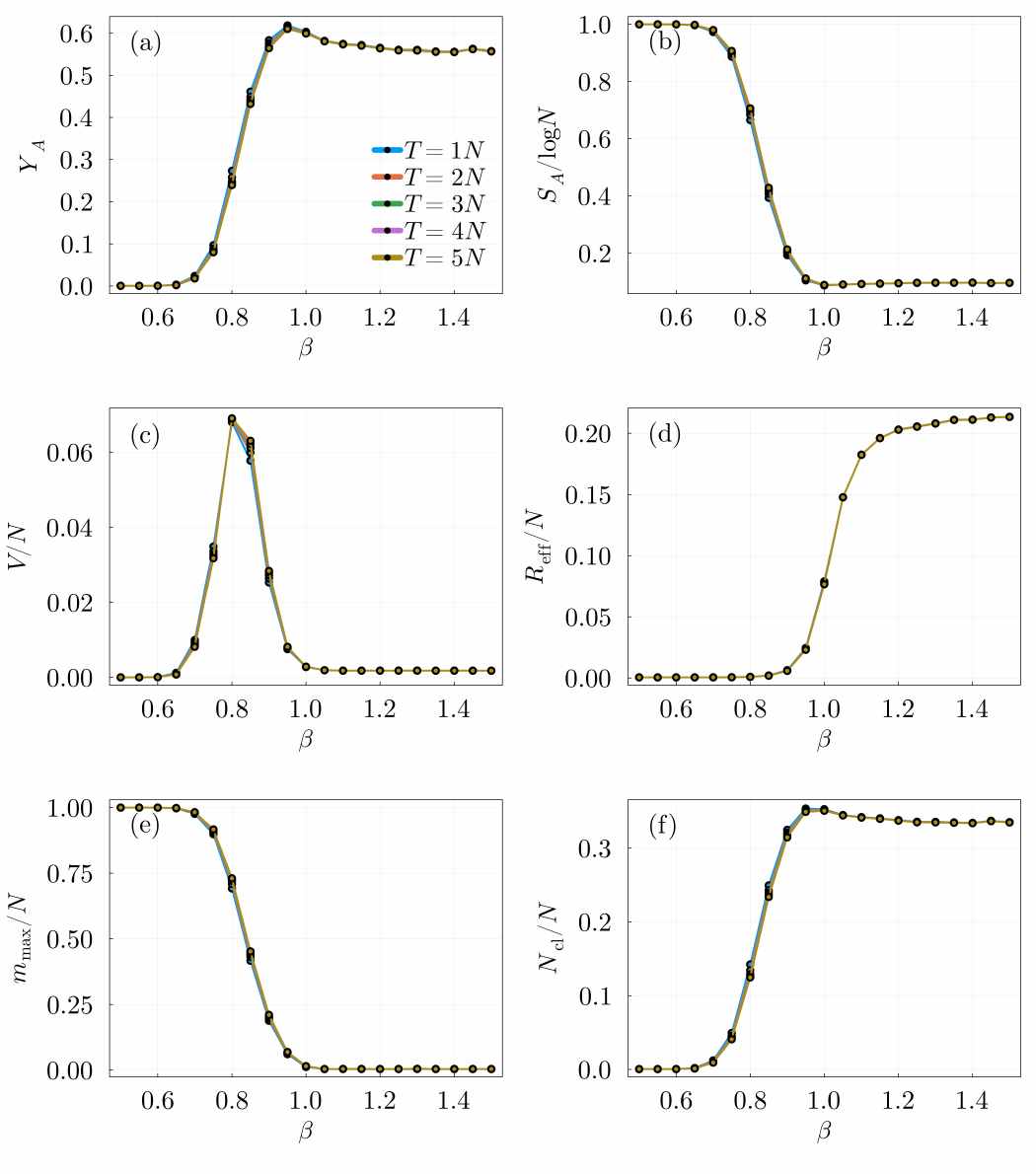}
    \caption{
    Observation-time dependence for Gaussian initial conditions at
    $N=2048$. The six panels show $Y_A$, $S_A/\log N$, $V/N$,
    $R_{\rm eff}/N$, $m_{\max}/N$, and $N_{\rm cl}/N$, measured at
    $T=N$, $2N$, $3N$, $4N$, and $5N$. 
    The curves at different observation times nearly coincide, showing
    that the six observables have reached a long-lived plateau by $T=N$
    over the time window studied.
    }
    \label{fig:supp_gaussian_time}
\end{figure*}

The configurations observed at $T=N$ are not strictly stationary at
finite $N$. Inter-cluster attention is exponentially small but remains
nonzero, allowing clusters to continue merging over longer time scales.
To display this slow finite-size evolution,
Fig.~\ref{fig:supp_gaussian_time} compares the same six observables at
\begin{equation}
    T=N,\ 2N,\ 3N,\ 4N,\ 5N
\end{equation}
for $N=2048$.

The six observables are nearly unchanged between
$T=N$ and $T=5N$. This agreement shows that the attention,
representation, and cluster observables have reached a long-lived
plateau by $T=N$ over the time window studied. We therefore use
$T=N$ as the common observation time for the finite-size comparison in
Fig.~\ref{fig:supp_gaussian_observables}. This finite-time convergence
does not imply that the configurations are exact stationary states at
finite $N$; exponentially weak inter-cluster attention may still
produce coarsening on much longer time scales.

\subsection{Nearly rank-collapsed initial conditions}
We next consider the nearly rank-collapsed initial state
\begin{equation}
    x_i(0)
    =
    \mathcal N_{\sqrt d}
    \left[
        \sqrt{1-\epsilon^2}\,u+\epsilon\,\eta_i
    \right]
\end{equation}
with $\epsilon=0.1$. The dynamics are organized by the effective
sharpness
\begin{equation}
    \alpha=\beta\epsilon^2,
\end{equation}
which controls the fluctuations of the tangent-space attention logits.

For this initial condition, we study
\begin{equation}
    Y_A,\qquad
    \frac{S_A}{\log N},\qquad
    V,\qquad
    R_{\rm eff},\qquad
    \frac{m_{\max}}{N},\qquad
    \frac{N_{\rm cl}}{N}.
\end{equation}
Unlike in the Gaussian case, $V$ and $R_{\rm eff}$ are not divided by
$N$. The tokens begin close to a common direction, and the descendant
clusters remain confined to the corresponding narrow cone. Their
global representation geometry therefore remains low rank, with
$V=O(1)$ and $R_{\rm eff}=O(1)$ even when the number of microscopic
clusters is extensive.

\begin{figure*}[t]
    \centering
    \includegraphics[width=\textwidth]
    {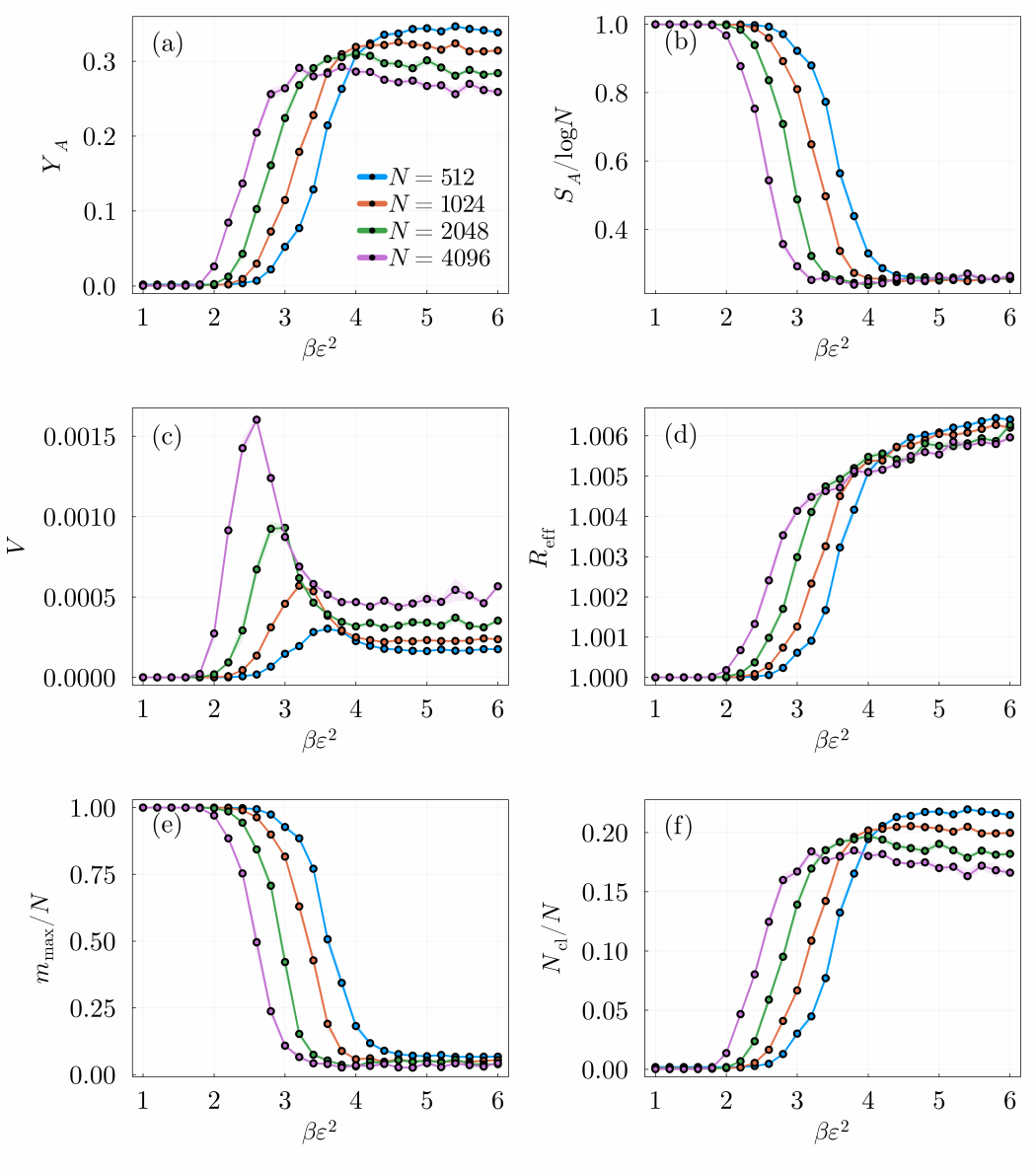}
    \caption{
    Attention, representation, and cluster observables for nearly
    rank-collapsed initial conditions with $\epsilon=0.1$, measured at
    $T=N$ for $N=512$, $1024$, $2048$, and $4096$.
    The six panels show $Y_A$, $S_A/\log N$, $V$, $R_{\rm eff}$,
    $m_{\max}/N$, and $N_{\rm cl}/N$ as functions of
    $\alpha=\beta\epsilon^2$. The attention and cluster observables
    show the fragmentation of the macroscopic parent cluster, while
    the small values of $V$ and $R_{\rm eff}$ show that the descendants
    remain confined to a low-rank narrow cone.
    }
    \label{fig:supp_collapsed_observables}
\end{figure*}

Figure~\ref{fig:supp_collapsed_observables} shows the six observables as
functions of $\alpha$ for
$N=512$, $1024$, $2048$, and $4096$, measured at $T=N$. At small
$\alpha$, broad averaging heals the initial angular width. The system
returns to a single aligned cluster, with $m_{\max}/N$ close to unity,
$N_{\rm cl}/N$ vanishing, $Y_A$ decreasing with system size, and
$S_A/\log N$ remaining close to unity.

As $\alpha$ increases, attention becomes localized and the
macroscopic parent cluster fragments. The attention and cluster
observables behave similarly to those obtained from Gaussian
initialization: $Y_A$ becomes finite, $S_A/\log N$ decreases,
$m_{\max}/N$ falls, and $N_{\rm cl}/N$ becomes finite. The essential
difference lies in the representation observables. Both $V$ and
$R_{\rm eff}$ remain small because the microscopic descendants retain
a strong common mode and occupy only a narrow angular cone. Thus,
attention condensation and extensive microscopic fragmentation do not
by themselves imply extensive representation rank.

\begin{figure*}[t]
    \centering
    \includegraphics[width=\textwidth]
    {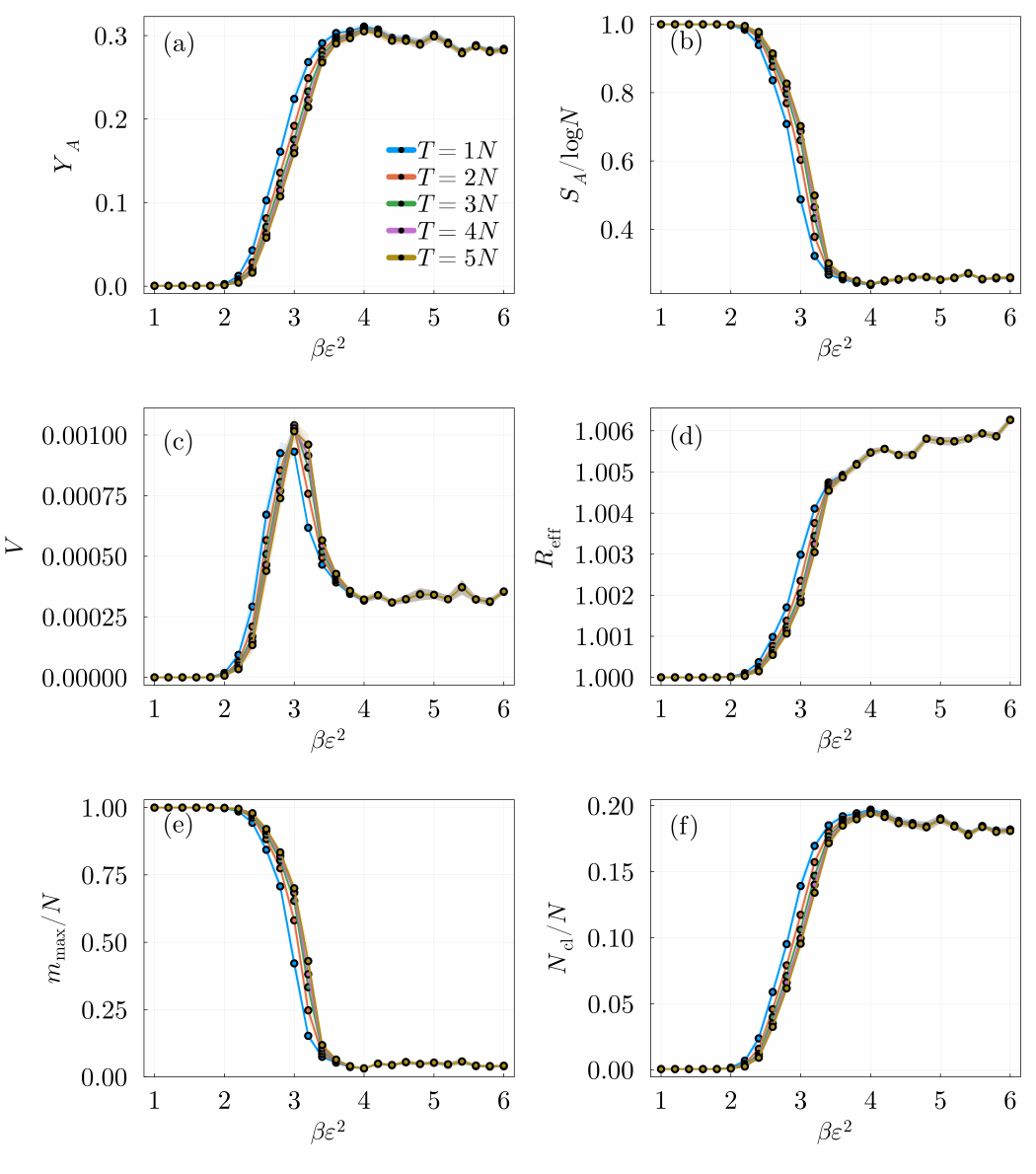}
    \caption{
    Observation-time dependence for nearly rank-collapsed initial
    conditions with $N=2048$ and $\epsilon=0.1$.
    The six panels show $Y_A$, $S_A/\log N$, $V$, $R_{\rm eff}$,
    $m_{\max}/N$, and $N_{\rm cl}/N$, measured at
    $T=N$, $2N$, $3N$, $4N$, and $5N$. The continued growth of the
    largest cluster and reduction in the number of clusters demonstrate
    slow finite-size coarsening, while the small values of $V$ and
    $R_{\rm eff}$ show that the evolving fragments remain confined to a
    common narrow cone.
    }
    \label{fig:supp_collapsed_time}
\end{figure*}

In contrast to the Gaussian case, the nearly rank-collapsed
configurations continue to evolve visibly over the accessible time
window. Figure~\ref{fig:supp_collapsed_time} compares the same
observables at
\begin{equation}
    T=N,\ 2N,\ 3N,\ 4N,\ 5N
\end{equation}
for $N=2048$ and $\epsilon=0.1$.

The largest cluster grows with increasing observation time, while the
total number of clusters decreases. The attention observables also
shift systematically, showing that the fragmented state has not yet
reached the time-independent plateau observed for Gaussian
initialization. These trends are consistent with slow finite-size coarsening, while the persistently small $V$ and $R_{\rm eff}$ indicate that the representation remains geometrically close to the rank-collapsed state.

At finite $N$, the narrow-cone fragmented state is
thus metastable rather than strictly stationary. Its coarsening time
grows exponentially with the overlap gap and with $\beta\sqrt N$, so
the fragmented state becomes stable when the thermodynamic limit is
taken before the long-time limit.

\bibliography{trajectory_attention_v2_refs}